\documentclass[twocolumn]{revtex4}

\usepackage{amsfonts}
\usepackage{amsmath}
\usepackage{amssymb}
\usepackage[dvips]{graphicx}
\usepackage{bm}

\newtheorem{theorem}{Theorem}

\begin{document}

\title{On Transgression Forms and Chern--Simons (Super)gravity}

\author{Fernando Izaurieta}
\email{fizaurie@theorie.physik.uni-muenchen.de}

\author{Eduardo Rodr\'{\i}guez}
\email{eduardo@theorie.physik.uni-muenchen.de}

\affiliation{Arnold Sommerfeld Center for Theoretical Physics,
Ludwig-Maximilians-Universit\"{a}t M\"{u}nchen, Theresienstra\ss e 37, D-80333
Munich, Germany}

\affiliation{Departament de F\'{\i}sica Te\`{o}rica, Universitat de Val\`{e}ncia, 46100
Burjassot, Val\`{e}ncia, Spain}

\affiliation{Departamento de F\'{\i}sica, Universidad de Concepci\'{o}n, Casilla 160-C,
Concepci\'{o}n, Chile}

\author{Patricio Salgado}
\email{pasalgad@udec.cl}

\affiliation{Departament de F\'{\i}sica Te\`{o}rica, Universitat de Val\`{e}ncia, 46100
Burjassot, Val\`{e}ncia, Spain}

\affiliation{Departamento de F\'{\i}sica, Universidad de Concepci\'{o}n, Casilla 160-C,
Concepci\'{o}n, Chile}

\preprint{LMU-ASC 77/05}

\date{December 1st, 2005}

\begin{abstract}
A transgression form is proposed as lagrangian for a gauge field theory. The
construction is first carried out for an arbitrary Lie Algebra $\mathfrak{g}$
and then specialized to some particular cases. We exhibit the action, discuss
its symmetries, write down the equations of motion and the boundary conditions
that follow from it, and finally compute conserved charges. We also present a
method, based on the iterative use of the Extended Cartan Homotopy Formula,
which allows one to (i) \emph{systematically} split the lagrangian in order to
appropriately reflect the subspaces structure of the gauge algebra, and (ii)
separate the lagrangian in bulk and boundary contributions. Chern--Simons Gravity
and Supergravity are then used as examples to illustrate the method. In the end
we discuss some further theoretical implications that arise naturally from
the mathematical structure being considered.
\end{abstract}

\maketitle



\section{\label{intro}Introduction}

The motivations for the study of higher-dimensional Gravity~\footnote{By
higher-dimensional we always mean, as in~\cite{Zum85}, \emph{in more than four
dimensions}.} have remained largely invariant since B. Zumino's paper
\textit{Gravity Theories in more than four dimensions}~\cite{Zum85}; we may
add here that String Theory (ST) has since grown into an all-encompassing
framework that guides and inspires research in high-energy physics. Among its
rich offspring, ST provides with a gravity lagrangian which includes, at
higher-order corrections in $\alpha^{\prime}$, higher powers of the curvature
tensor. The potential incompatibilities between the ghost particles usually
associated with these terms and the ghost-free ST were first analyzed in
\cite{Zwi85}, where it was pointed out that a proper combination of
curvature-squared terms leads to `ghost-free, non-trivial gravitational
interactions for dimensions higher than four'. The key point, already
conjectured by Zwiebach, and confirmed later by Zumino, is that the allowed
terms in $d$ dimensions are the \emph{dimensional continuations} of all the
Euler densities of dimension lower than $d$. For even $d$, one has the
seemingly odd choice of also adding the Euler density corresponding to $d $,
which, being a total derivative, does not contribute to the equations of
motion. However, this term is crucial in order to attain a proper
regularization for the conserved charges (like mass and angular momentum)
\cite{Zan99,Zan00}. Interestingly enough, the ghost-free,
higher-power-in-curvature lagrangians considered by Zumino had been introduced
much earlier, in a completely classical context, by D.~Lovelock~\cite{Lov70}
(it is also noteworthy the contribution of C.~Lanczos, Ref.~\cite{Lan38}).

Lanczos--Lovelock (LL) lagrangians have been extensively studied (for some
recent work, check, e.g.,
\cite{Des05,Zeg05,Hen05,Wil04,Aie05,Iza04,All03,Sal03,Aro00,Cri00}). There are
typically~\footnote{Here $\left[  x\right]  $ denotes the integer part of
$x$.} $1+\left[  d/2\right]  $ ghost-free LL lagrangians in $d$ dimensions,
which can be linearly combined, with arbitrary coefficients $\alpha_{p}$,
$p=0,1,\ldots,\left[  d/2\right]  $, into a full-fledged gravity lagrangian.
The first term in the series, without any curvature contributions $\left(
p=0\right)  $, corresponds to a cosmological term, while the second one,
linear in curvature $\left(  p=1\right)  $, is nothing else than the usual
Einstein--Hilbert (EH) lagrangian.

The next major step in our programme is the embedding of the full LL
lagrangian into the broader scheme of Chern--Simons (CS) theory. As first
shown by Chamseddine~\cite{Cha89,Cha90} and Ba\~{n}ados, Teitelboim and
Zanelli~\cite{Ban93}, there is a special choice of the $\alpha_{p}$
coefficients that brings the LL lagrangian in odd dimensions only a total
derivative away from a CS lagrangian. To be more explicit, take $\bm{A}$ to be
a $\mathfrak{so}\left(  d+1\right)  $-valued, one-form gauge
connection~\footnote{This could be the de Sitter Algebra $\mathfrak{so}\left(
d,1\right)  $ or the anti-de Sitter Algebra $\mathfrak{so}\left(
d-1,2\right)  $; we are not making the distinction here.} and pick the
Levi-Civita symbol $\varepsilon$ as an invariant tensor. The CS lagrangian
built with these two ingredients is then equal to a LL lagrangian (with the
chosen coefficients) \emph{plus a total derivative}.

An alternative way of deriving these `canonical' coefficients rests on the
physical requirement of well-defined dynamics~\cite{Tro99}. This argument can
also be made in even dimensions, where the resulting lagrangian has a
Born--Infeld-like form, leading to the associated theory being dubbed
`Born--Infeld (BI) Gravity'. Conserved charges for even-dimensional BI gravity
can be computed via a direct application of Noether's Theorem
\cite{Zan99,Zan00}. The formula so obtained correctly reproduces mass and
angular momentum for a host of exact solutions, without requiring
regularization or the subtraction of an ad-hoc background. This success is
however harder to replicate for their odd-dimensional counterpart, CS gravity.
Na\"{\i}ve application of Noether's Theorem yields in this case a formula for
the charges that fails to give the physically correct values for at least one
solution. A beautiful resolution of this uncomfortable situation has been
recently proposed by Mora, Olea, Troncoso and Zanelli in~\cite{Mor04,Ole04}.
They add a carefully selected boundary term to the CS action which renders it
both finite and capable of producing well-defined charges. Although this
boundary term is deduced from purely gravitational arguments, the last
paragraph in~\cite{Mor04} already points to the ultimate reason for the
action's remarkable properties: it can be regarded as a \emph{transgression
form}.

Transgression forms are the matrix where CS forms stem from~\cite{Nak03,Azc95}%
. In this paper we shall be mainly concerned with the formulation of what may
be called Transgression Gauge Field Theory (TGFT); i.e., a classical, gauge
field theory whose lagrangian is a transgression form.

The organization of the paper goes as follows. In section~\ref{cs} we briefly
review CS Theory and comment on its relation to the Chern--Weil theorem.
Section~\ref{tgft} brings in the transgression form as a lagrangian for a
gauge field theory. We discuss its symmetries, write down the equations of
motion and the boundary conditions that follow from it, and finally compute
conserved charges. LL Gravity is recovered as a first example in section
\ref{llg}. The Extended Cartan Homotopy Formula (ECHF) is presented in section
\ref{echf}, and in section~\ref{trac} it is shown to be an extremely useful
tool in formulating gravity and supergravity. Finally, important theoretical
issues are given some thought in section~\ref{sur}.

\section{\label{cs}Chern--Simons Theory: A Review}

Let $\mathfrak{g}$\ be a Lie algebra over some field. Essential objects in
everything that follows will be $\mathfrak{g}$-valued differential forms on
some space-time manifold $M$, which we shall denote by italic boldface. When
$\bm{P}$ is a $p$-form and $\bm{Q}$ is a $q$-form, then its Lie bracket
$\left[  \cdot,\cdot\right]  $ has the following symmetry:
\begin{equation}
\left[  \bm{P},\bm{Q}\right]  =\left(  -1\right)  ^{pq}\left[
\bm{Q},\bm{P}\right]  .
\end{equation}
Given the Lie bracket and a one-form $\bm{A}$ it is always possible to define
a `covariant derivative' D as
\begin{equation}
\text{D}\bm{Z}\equiv\text{d}\bm{Z}+\left[  \bm{A},\bm{Z}\right]  ,
\end{equation}
where d denotes the usual exterior derivative. This covariant derivative has
the (defining) property that, if $\bm{Z}$ transforms as a tensor and $\bm{A}$
as a connection under $\mathfrak{g}$, then D$\bm{Z}$ will also transform as a tensor.

Let $\left\langle \cdots\right\rangle _{r}$ denote a $\mathfrak{g}$-invariant
symmetric polynomial
\begin{equation}
\left\langle \cdots\right\rangle _{r}:\;\underset{r}{\underbrace
{\mathfrak{g\times\cdots\times g}}}\;\rightarrow\mathbb{C}%
\end{equation}
of some fixed rank $r$. The invariance requirement for $\left\langle
\cdots\right\rangle _{r}$ essentially boils down to~\footnote{The {\small LHS}
of (\ref{dzr}) is to be understood as defined through Leibniz's rule, i.e.,
$\left\langle \text{D}\left(  \bm{Z}_{1}\cdots\bm{Z}_{r}\right)  \right\rangle
_{r}$ actually stands for $\left\langle \left(  \text{D}\bm{Z}_{1}\right)
\bm{Z}_{2}\cdots\bm{Z}_{r}\right\rangle _{r}+\left(  -1\right)  ^{p_{1}%
}\left\langle \bm{Z}_{1}\left(  \text{D}\bm{Z}_{2}\right)  \bm{Z}_{3}%
\cdots\bm{Z}_{r}\right\rangle _{r}+\ldots+\left(  -1\right)  ^{p_{1}%
+\ldots+p_{r-1}}\left\langle \bm{Z}_{1}\cdots\bm{Z}_{r-1}\left(
\text{D}\bm{Z}_{r}\right)  \right\rangle _{r}$, where $p_{k}$ denotes the rank
of $\bm{Z}_{k}$.}
\begin{equation}
\left\langle \text{D}\left(  \bm{Z}_{1}\cdots\bm{Z}_{r}\right)  \right\rangle
_{r}=\text{d}\left\langle \bm{Z}_{1}\cdots\bm{Z}_{r}\right\rangle
_{r},\label{dzr}%
\end{equation}
where $\left\{  \bm{Z}_{k},k=1,\ldots,r\right\}  $ is a set of $\mathfrak{g}%
$-valued differential forms. The symmetry requirement for $\left\langle
\cdots\right\rangle _{r}$ implies that, for any $p$-form $\bm{P}$ and $q$-form
$\bm{Q}$, we have
\begin{equation}
\left\langle \cdots\bm{PQ}\cdots\right\rangle _{r}=\left(  -1\right)
^{pq}\left\langle \cdots\bm{QP}\cdots\right\rangle _{r}.
\end{equation}
This remains valid even in the case of $\mathfrak{g}$ being a superalgebra,
due to the Grassmann nature of the parameters that multiply fermionic generators.

In what follows we shall usually drop the subscript $r$ in the invariant
polynomial, as we will only use one fixed rank (to be specified below).

A CS lagrangian in $d=2n+1$ dimensions is defined to be the following local
function~\footnote{It should be noted that, since $\bm{A}$ is a one-form,
$\bm{A}^{2}=\frac{1}{2}\left[  \bm{A},\bm{A}\right]  $ is a $\mathfrak{g}%
$-valued two-form (as is $\text{d}\bm{A}$).} of a one-form gauge connection
$\bm{A}$:
\begin{equation}
L_{\text{CS}}^{\left(  2n+1\right)  }\left(  \bm{A}\right)  =\left(
n+1\right)  k\int_{0}^{1}dt\left\langle \bm{A}\left(  t\text{d}\bm{A}+t^{2}%
\bm{A}^{2}\right)  ^{n}\right\rangle ,\label{lcs}%
\end{equation}
where $\left\langle \cdots\right\rangle $ denotes a $\mathfrak{g}$-invariant
symmetric polynomial of rank $r=n+1$ and $k$ is a constant. One important fact
to note here is that CS forms are only locally defined. To see this, we have
to consider the following result~\cite{Nak03,Azc95}:

\begin{theorem}
(Chern--Weil). Let $\bm{A}$ and $\bar{\bm{A}}$ be two one-form gauge
connections on a fiber bundle over a $\left(  2n+1\right)  $-dimensional
manifold $M$, and let $\bm{F}$ and $\bar{\bm{F}}$ be the corresponding
curvatures. Then, with the above notation,
\begin{equation}
\left\langle \bm{F}^{n+1}\right\rangle -\left\langle \bar{\bm{F}}%
^{n+1}\right\rangle =\text{d}Q_{\bm{A}\leftarrow\bar{\bm{A}}}^{\left(
2n+1\right)  },\label{cwt}%
\end{equation}
where
\begin{equation}
Q_{\bm{A}\leftarrow\bar{\bm{A}}}^{\left(  2n+1\right)  }\equiv\left(
n+1\right)  \int_{0}^{1}dt\left\langle \bm{\theta F}_{t}^{n}\right\rangle
\label{tra}%
\end{equation}
is called a \textbf{transgression form} and we have defined
\begin{align}
\bm{\theta}  &  \equiv\bm{A}-\bar{\bm{A}},\label{thk}\\
\bm{A}_{t}  &  \equiv\bar{\bm{A}}+t\bm{\theta },\label{at}\\
\bm{F}_{t}  &  \equiv\text{d}\bm{A}_{t}+\bm{A}_{t}^{2}.\label{ft}%
\end{align}

\end{theorem}

A sketch of the proof is given in Appendix~\ref{prcw} in order to get an
intuitive, `physical' sense of the theorem's content. The theorem is also
deduced as a corollary of the ECHF in section~\ref{cwt1}.

Setting $\bar{\bm{A}}=0$ in (\ref{tra}) gives the CS form as
\begin{align}
Q^{\left(  2n+1\right)  }\left(  \bm{A}\right)   &  =Q_{\bm{A}\leftarrow
0}^{\left(  2n+1\right)  },\nonumber\\
&  =\left(  n+1\right)  \int_{0}^{1}dt\left\langle \bm{A}\left(
t\text{d}\bm{A}+t^{2}\bm{A}^{2}\right)  ^{n}\right\rangle .
\end{align}
The Chern--Weil theorem for this particular case shows that d$\delta
Q^{\left(  2n+1\right)  }\left(  \bm{A}\right)  =0$; this implies that $\delta
Q^{\left(  2n+1\right)  }\left(  \bm{A}\right)  $ may be locally written as
d$\Omega$ for some $\Omega$ under gauge transformations. But since a
connection cannot be globally set to zero unless the bundle (topology) is
trivial, CS forms turn out to be only locally defined. Of course, this is only
a problem if one insists on using a CS form as a lagrangian, since then one
has to integrate it over all of $M$ to get the action. Nevertheless, CS forms
are used as lagrangians mainly because (i) they lead to gauge theories with a
fiber-bundle structure, whose only dynamical field is a one-form gauge
connection $\bm{A}$ and (ii) they do change by only a total derivative under
gauge transformations. When we choose $\mathfrak{g}=\mathfrak{so}\left(
2n+2\right)  $ and write $\bm{A}$ as%
\begin{equation}
\bm{A}=e^{a}\bm{P}_{a}+\frac{1}{2}\omega^{ab}\bm{J}_{ab},
\end{equation}
then the CS form provides with a \emph{background-free} gravity theory (since
metricity is given by the vielbein, which is just one component of $\bm{A}$).
An explicit realization of $d=11$ Supergravity (SUGRA) in terms of a CS
lagrangian has even shed some new light on the old problem of why our world
seems to be four-dimensional~\cite{Has03,Has05}.

On the other hand, a transgression form is in principle globally well defined
and also, as we will see, invariant under gauge transformations. We now turn
to the discussion of transgression forms used as lagrangians for gauge field
theories --- TGFT.

\section{\label{tgft}The Transgression Form as a Lagrangian}

\subsection{TGFT Field Equations}

We consider a gauge theory on an orientable $\left(  2n+1\right)
$-dimensional manifold $M$ defined by the action
\begin{align}
S_{\text{T}}^{\left(  2n+1\right)  }\left[  \bm{A},\bar{\bm {A}}\right]   &
=k\int_{M}Q_{\bm{A}\leftarrow\bar{\bm{A}}}^{\left(  2n+1\right)  }\nonumber\\
&  = \left(  n+1\right)  k \int_{M}\int_{0}^{1}dt\left\langle
\bm{\theta F}_{t}^{n}\right\rangle ,\label{st}%
\end{align}
where $k$ is a constant (see previous section for notation and conventions).
Eq.~(\ref{st}) describes (the dynamics of) a theory with two independent
fields, namely the two one-form gauge connections $\bm{A}$ and $\bar{\bm{A}}$.
These fields enter the action in a rather symmetrical way; for instance,
interchange of both connections produces a sign difference, i.e.,
\begin{equation}
S_{\text{T}}^{\left(  2n+1\right)  }\left[  \bar{\bm{A}},\bm{A}\right]
=-S_{\text{T}}^{\left(  2n+1\right)  }\left[  \bm{A},\bar{\bm{A}}\right]  .
\end{equation}

A striking new feature of the action (\ref{st}) is the presence of two
independent one-form gauge connections, $\bm{A}$ and $\bar{\bm
{A}}$. The physical interpretation of this will be made clear in the next
sections through several examples, which we discuss in some detail.

Performing independent variations of $\bm{A}$ and $\bar{\bm {A}}$ in
(\ref{st}) we get after some algebra (see Appendix~\ref{feqbc})
\begin{equation}
\delta S_{\text{T}}^{\left(  2n+1\right)  }=\left(  n+1\right)  k\int
_{M}\left(  \left\langle \delta\bm{AF}^{n}\right\rangle -\left\langle
\delta\bar{\bm{A}}\bar{\bm{F}}^{n}\right\rangle \right)  +\int_{\partial
M}\Theta,\label{var}%
\end{equation}
with
\begin{equation}
\Theta=n\left(  n+1\right)  k\int_{0}^{1}dt\left\langle \delta\bm{A}_{t}%
\bm{\theta F}_{t}^{n-1}\right\rangle .\label{th}%
\end{equation}
This result (but for the exact form of the boundary term $\Theta$) may be
readily guessed just by looking at the Chern--Weil theorem. An explicit
computation gives us (\ref{th}).

The TGFT field equations can be directly read off from the variation
(\ref{var}). They are
\begin{align}
\left\langle \bm{F}^{n}\bm{G}_{a}\right\rangle  &  =0,\label{fg1}\\
\left\langle \bar{\bm{F}}^{n}\bm{G}_{a}\right\rangle  &  =0,\label{fg2}%
\end{align}
where $\left\{  \bm{G}_{a},a=1,\ldots,\dim\left(  \mathfrak{g}\right)
\right\}  $ is a basis for $\mathfrak{g}$. Boundary conditions are obtained by
demanding the vanishing of $\Theta$ on $\partial M$:
\begin{equation}
\left.  \int_{0}^{1}dt\left\langle \delta\bm{A}_{t}\bm{\theta F}_{t}%
^{n-1}\right\rangle \right\vert _{\partial M}=0.\label{bc}%
\end{equation}
Eqs.~(\ref{fg1})--(\ref{bc}) tell us about two independent CS theories living
on a manifold $M$ which are inextricably linked at the boundary. Below we
shall see some examples where this story is told, albeit in a rather simple
and surprising way.

\subsection{\label{sym}Symmetries}

There are two major independent symmetries lurking in our TGFT action, eq.
(\ref{st}). The first of them is a built-in symmetry, guaranteed from the
outset by our use of differential forms throughout: it is diffeomorphism
invariance. Although straightforward, the symmetry is far-reaching, as is
proved by the fact that it leads to non-trivial conserved charges.

The second symmetry is gauge symmetry. Under a continuous, local gauge
transformation generated by a group element $g=\exp\left(  \lambda
^{a}\bm{G}_{a}\right)  $, the connections change as
\begin{align}
\bm{A}  &  \rightarrow\bm{A}^{\prime}=g\left(  \bm{A}-g^{-1}\text{d}g\right)
g^{-1},\label{ag}\\
\bar{\bm{A}}  &  \rightarrow\bar{\bm{A}}^{\prime}=g\left(  \bar{\bm{A}}%
-g^{-1}\text{d}g\right)  g^{-1}.\label{abg}%
\end{align}
Invariance of the TGFT lagrangian under (\ref{ag})--(\ref{abg}) is guaranteed
by (i) the fact that both $\bm{\theta}$ and $\bm{F}_{t}$ transform as tensors
and (ii) the invariant nature of the symmetric polynomial $\left\langle
\cdots\right\rangle $.

Crucially, and in stark contrast with the CS case, there are no boundary terms
left after the gauge transformation: the TGFT action (\ref{st}) is fully
invariant rather than pseudo-invariant. A pseudo-invariant lagrangian is one
that changes by a closed form under gauge transformations. In this case the
lagrangian ceases to be univocally defined; quite naturally, the addition
of an \emph{arbitrary} exact form cannot be ruled out on the grounds of
symmetry alone. A direct consequence is an inherent ambiguity in the boundary
conditions that must follow from the action. Any conserved
charges computed from a pseudo-invariant lagrangian will also be changed by this
modification, rendering them also ambiguous. A TGFT suffers from none of
this problems; we cannot add an arbitrary closed form to the
Lagrangian, since that would destroy the symmetry. This means that the
charges and boundary conditions derived from the TGFT action are in principle
physically meaningful.

\subsection{\label{carg}Conserved Charges}

In order to fix the notation and conventions, we briefly review here Noether's
Theorem in the language of differential forms~\cite{Zan99,Zan00}. Let
$L\left(  \varphi\right)  $ be a lagrangian $d$-form for some set of fields
$\varphi$. An infinitesimal functional variation $\delta\varphi$ induces an
infinitesimal variation $\delta L$,
\begin{equation}
\delta L=E\left(  \varphi\right)  \delta\varphi+\text{d}\Theta\left(
\varphi,\delta\varphi\right)  ,
\end{equation}
where $E\left(  \varphi\right)  =0$ are the equations of motion and $\Theta$
is a boundary term which depends on $\varphi$ and its variation $\delta
\varphi$. When $\delta\varphi$ corresponds to a gauge transformation, the
off-shell variation of $L$ equals (at most) a total derivative, $\delta L=$
d$\Omega$. Noether's Theorem then states that the current~\footnote{We have
chosen to write the $\left(  d-1\right)  $-form Noether current as the Hodge
$\star$-dual of a one-form $J$, but this is in no way mandatory.}
\begin{equation}
\left.  \star J_{\text{gauge}}\right.  =\Omega-\Theta\left(  \varphi
,\delta_{\text{gauge}}\varphi\right) \label{jg}%
\end{equation}
is on-shell conserved; i.e., d$\left.  \star J_{\text{gauge}}\right.  =0$. An
analogous statement is valid for a diffeomorphism generated by a vector field
$\xi$, $\delta x^{\mu}=\xi^{\mu}\left(  x\right)  $. In this case the
conserved current has the form~\footnote{Here I$_{\xi}$ is the
\textit{contraction operator} (also called interior product and denoted as
$\xi\rfloor$), which takes a $p$-form $\alpha=\left(  1/p!\right)  \alpha
_{\mu_{1}\cdots\mu_{p}}$d$x^{\mu_{1}}\cdots$d$x^{\mu_{p}}$ into the $\left(
p-1\right)  $-form I$_{\xi}\alpha=\left(  1/\left(  p-1\right)  !\right)
\xi^{\mu_{1}}\alpha_{\mu_{1}\cdots\mu_{p}}$d$x^{\mu_{2}}\cdots$d$x^{\mu_{p}}$.
The Lie derivative $\pounds _{\xi}$ may be written in terms of I$_{\xi}$ and
the exterior derivative d as $\pounds _{\xi}= \text{dI}_{\xi}+\text{I}_{\xi}%
$d.}
\begin{equation}
\left.  \star J_{\text{diff}}\right.  =-\Theta\left(  \varphi,\delta
_{\text{diff}}\varphi\right)  -\text{I}_{\xi}L.\label{jd}%
\end{equation}
In eqs.~(\ref{jg}) and (\ref{jd}) it is understood that we replace in $\Theta$
the variation of $\varphi$ corresponding to the gauge transformation or the
diffeomorphism, respectively.

When formul\ae \ (\ref{jg})--(\ref{jd}) are applied to the TGFT lagrangian
[cf. eq.~(\ref{st})]%
\begin{equation}
L_{\text{T}}^{\left(  2n+1\right)  }=\left(  n+1\right)  k \int_{0}%
^{1}dt\left\langle \bm{\theta F}_{t}^{n}\right\rangle ,
\end{equation}
the following conserved currents are obtained (see Appendix~\ref{ccs} for a
derivation):
\begin{align}
\left.  \star J_{\text{gauge}}\right.   &  =n\left(  n+1\right)  k\text{d}%
\int_{0}^{1}dt\left\langle \bm{\lambda\theta F}_{t}^{n-1}\right\rangle
,\label{jgt}\\
\left.  \star J_{\text{diff}}\right.   &  =n\left(  n+1\right)  k\text{d}%
\int_{0}^{1}dt\left\langle \text{I}_{\xi}\bm{A}_{t}\bm{\theta F}_{t}%
^{n-1}\right\rangle .\label{jdt}%
\end{align}
Here $\bm{\lambda}$ is a local, $\mathfrak{g}$-valued 0-form parameter that
defines an infinitesimal gauge transformation via $\delta\bm{A}=-$%
D$_{\bm{A}}\bm{\lambda}$, $\delta\bar{\bm{A}}=-$D$_{\bar{\bm{A}}}\bm{\lambda}$
and $\xi$ is a vector field that generates an infinitesimal diffeomorphism,
which acts on $\bm{A}$ and $\bar{\bm{A}}$ as $\delta\bm{A}=-\pounds _{\xi
}\bm{A}$, $\delta\bar{\bm{A}}=-\pounds _{\xi}\bar{\bm{A}}$. In writing
(\ref{jgt}) and (\ref{jdt}) we have dropped terms proportional to the
equations of motion, so that the currents are only defined on-shell. This
allows writing them as total derivatives, a step which renders verification of
the conservation law d$\left.  \star J\right.  =0$ trivial.

Assuming the space-time manifold $M$ to have the topology $M=\mathbb{R}%
\times\Sigma$, we can integrate (\ref{jgt}) and (\ref{jdt}) over the `spatial
section' $\Sigma$ to get the conserved charges
\begin{align}
Q_{\text{gauge}}  &  =n\left(  n+1\right)  k\int_{\partial\Sigma}\int_{0}%
^{1}dt\left\langle \bm{\lambda\theta F}_{t}^{n-1}\right\rangle ,\label{qgt}\\
Q_{\text{diff}}  &  =n\left(  n+1\right)  k\int_{\partial\Sigma}\int_{0}%
^{1}dt\left\langle \text{I}_{\xi}\bm{A}_{t}\bm{\theta F}_{t}^{n-1}%
\right\rangle .\label{qdt}%
\end{align}
Stokes' Theorem allows us to restrict the integration to the \emph{boundary}
of the spatial section $\Sigma$.

To summarize, we have given explicit formul\ae \ to compute conserved charges
for the TGFT lagrangian as integrals over the boundary of an spatial section
in space-time. The general proof of finiteness for these charges remains as an
open problem; however, examples already exist where this is explicitly
confirmed~\cite{Mor04}.

The charges (\ref{qgt}) and (\ref{qdt}) are trivially invariant under
diffeomorphisms, since they're built out of differential forms. Invariance
under gauge transformations is slightly less straightforward; under
$\delta\bm{A}=-$D$\bm{\lambda}$, $Q_{\text{gauge}}$ remains invariant and
$Q_{\text{diff}}$ transforms as
\begin{equation}
\delta_{\bm{\lambda}}Q_{\text{diff}}=-n\left(  n+1\right)  k\int
_{\partial\Sigma}\int_{0}^{1}dt\left\langle \pounds _{\xi}%
\bm{\lambda\theta F}_{t}^{n-1}\right\rangle .\label{dqdt}%
\end{equation}
From (\ref{dqdt}) we see that a sufficient condition to ensure invariance of
$Q_{\text{diff}}$ under gauge transformations is to demand the transformation
to satisfy $\pounds _{\xi}\bm{\lambda}=0$ on $\partial\Sigma$. That is,
$Q_{\text{diff}}$ is invariant under those restricted gauge transformations
that fulfill this condition.

As a final remark on the charge formul\ae \ (\ref{qgt})--(\ref{qdt}), we would
like to point out that both $Q_{\text{gauge}}$ and $Q_{\text{diff}}$ flip
signs under the interchange $\bm{A}\leftrightarrows\bar{\bm {A}}$; we may then
interpret this operation as `charge conjugation'.

\section{\label{llg}An Example: LL Gravity as a TGFT}

As already mentioned in the Introduction, the lagrangian for odd-dimensional
LL gravity with the canonical coefficients is only a total derivative away
from a CS form. To see this, consider for definiteness the AdS Algebra in
$d=2n+1$ dimensions,
\begin{align}
\left[  \bm{P}_{a},\bm{P}_{b}\right]   &  =\frac{1}{\ell^{2}}\bm{J}_{ab}%
,\label{PP}\\
\left[  \bm{J}_{ab},\bm{P}_{c}\right]   &  =\eta_{cb}\bm{P}_{a}-\eta
_{ca}\bm{P}_{b},\label{JP}\\
\left[  \bm{J}_{ab},\bm{J}_{cd}\right]   &  =\eta_{cb}\bm{J}_{ad}-\eta
_{ca}\bm{J}_{bd}+\eta_{db}\bm{J}_{ca}-\eta_{da}\bm{J}_{cb},\label{JJ0}%
\end{align}
where $\ell$ is a length (the AdS radius). The $\mathfrak{so}\left(
2n,2\right)  $-valued one-form gauge connection has the form
\begin{equation}
\bm{A}=e^{a}\bm{P}_{a}+\frac{1}{2}\omega^{ab}\bm{J}_{ab},\label{AAdS}%
\end{equation}
where we identify $e^{a}$ with the vielbein and $\omega^{ab}$ with the spin connection.

There are several choices one can make for an invariant polynomial; perhaps
the simplest of them is the one for which%
\begin{equation}
\left\langle \bm{J}_{a_{1}a_{2}}\cdots\bm{J}_{a_{2n-1}a_{2n}}\bm{P}_{a_{2n+1}%
}\right\rangle =\frac{2^{n}}{\ell}\varepsilon_{a_{1}\cdots a_{2n+1}%
},\label{eps}%
\end{equation}
with all other combinations vanishing.

When we use the connection (\ref{AAdS}) and the invariant polynomial
(\ref{eps}) in the general formula for the CS form, eq.~(\ref{lcs}), we get
the LL lagrangian with the canonical coefficients plus a total derivative.
What we would like to point out here is that this LL lagrangian, without the
total derivative coming from the CS form, may also be regarded as a
Transgression form. As a matter of fact, when we choose%
\begin{align}
\bm{A}  &  =\frac{1}{2}\omega^{ab}\bm{J}_{ab}+e^{a}\bm{P}_{a},\\
\bar{\bm{A}}  &  =\frac{1}{2}\omega^{ab}\bm{J}_{ab},
\end{align}
and the same invariant tensor (\ref{eps}), then the TGFT lagrangian [cf. eq.
(\ref{st})] reads \cite{Cha89,Cha90}
\begin{align}
L_{\bm{\omega}+\bm{e}\leftarrow\bm{\omega}}^{\left(  2n+1\right)  }  &
=\left(  n+1\right)  \frac{k}{\ell}\varepsilon_{a_{1}\cdots a_{2n+1}}%
\times\nonumber\\
&  \times\int_{0}^{1}dtF_{t}^{a_{1}a_{2}}\cdots F_{t}^{a_{2n-1}a_{2n}%
}e^{a_{2n+1}},\label{lcst}%
\end{align}
where
\begin{equation}
F_{t}^{ab}=R^{ab}+\frac{t^{2}}{\ell^{2}}e^{a}e^{b}.
\end{equation}
From eq.~(\ref{lcst}) we learn that the choice (\ref{eps}) of invariant
symmetric polynomial effectively amounts to excluding the torsion from
appearing explicitly in the lagrangian (although it is not assumed to vanish).
It is interesting to note that the $t$-integration in (\ref{lcst}) manages to
exactly reproduce the canonical coefficients for the LL polynomial
\cite{Ban93}.

It is not at all obvious that the TGFT lagrangian (\ref{lcst}) and the CS
lagrangian (\ref{lcs}) should differ only by a total derivative. The fact that
they do is associated with the particular form of the invariant polynomial
used in both cases, namely the Levi-Civita tensor. In fact, let us recall that
the CS lagrangian locally satisfies
\begin{equation}
\text{d}L_{\text{CS}}^{\left(  2n+1\right)  }=k\left\langle \bm{F}^{n+1}%
\right\rangle ,
\end{equation}
whereas the TFGT lagrangian satisfies
\begin{equation}
\text{d}L_{\bm{\omega}+\bm{e}\leftarrow\bm{\omega}}^{\left(  2n+1\right)
}=k\left(  \left\langle \bm{F}^{n+1}\right\rangle -\left\langle \bar
{\bm{F}}^{n+1}\right\rangle \right)  .
\end{equation}
Our choice for the invariant polynomial now implies that
\[
\left\langle \bar{\bm{F}}^{n+1}\right\rangle =0,
\]
and we see that both lagrangians can only differ the way they do. This kind of
structure for the invariant polynomial $\left\langle \cdots\right\rangle $
will have important consequences, as we will see in the next sections.

\section{\label{echf}The Extended Cartan Homotopy Formula}

In principle, the TGFT lagrangian in its full generality [cf. eq.
(\ref{st})],
\begin{equation}
L_{\bm{A}\leftarrow\bar{\bm{A}}}^{\left(  2n+1\right)  }=\left(  n+1\right)
k\int_{0}^{1}dt\left\langle \bm{\theta F}_{t}^{n}\right\rangle ,\label{lt}%
\end{equation}
has all the information one needs about the theory; as shown in section
\ref{tgft}, it is possible to write down general expressions for the equations
of motion, the boundary conditions and the conserved charges without ever
bothering to say what the gauge group is supposed to be. In practice, however,
one often deals with a fixed gauge group or supergroup with several distinct
subgroups which have an individual, clear physical meaning. It would then be
desirable to have a systematic procedure to split the lagrangian (\ref{lt})
into pieces that reflect this group structure.

In this section we discuss a tool on which a separation method can be built
(see section \ref{trac}). Let us begin by noticing that,
according to the Chern--Weil Theorem (see section~\ref{cs}), the following
combination of derivatives of transgression forms identically vanishes:
\begin{align}
&  \text{d}Q_{\bm{A}\leftarrow\bar{\bm{A}}}^{\left(  2n+1\right)  }%
+\text{d}Q_{\tilde{\bm{A}}\leftarrow\bm{A}}^{\left(  2n+1\right)  }%
+\text{d}Q_{\bar{\bm{A}}\leftarrow\tilde{\bm{A}}}^{\left(  2n+1\right)
}\nonumber\\
&  =\left\langle \bm{F}^{n+1}\right\rangle -\left\langle \bar{\bm{F}}%
^{n+1}\right\rangle +\nonumber\\
&  +\left\langle \tilde{\bm{F}}^{n+1}\right\rangle -\left\langle
\bm{F}^{n+1}\right\rangle +\\
&  +\left\langle \bar{\bm{F}}^{n+1}\right\rangle -\left\langle \tilde
{\bm{F}}^{n+1}\right\rangle \\
&  =0.
\end{align}
Here $\bm{A}$, $\tilde{\bm{A}}$ and $\bar{\bm{A}}$ are three arbitrary,
one-form gauge connections, with $\bm{F}$, $\tilde{\bm{F}}$ and $\bar{\bm{F}}
$ being the corresponding curvatures. This vanishing further implies that we
can (at least locally) write
\begin{equation}
Q_{\bm{A}\leftarrow\bar{\bm{A}}}^{\left(  2n+1\right)  }+Q_{\tilde
{\bm{A}}\leftarrow\bm{A}}^{\left(  2n+1\right)  }+Q_{\bar{\bm{A}}%
\leftarrow\tilde{\bm{A}}}^{\left(  2n+1\right)  }=\text{d}Q_{\bm{A}\leftarrow
\tilde{\bm{A}}\leftarrow\bar{\bm{A}}}^{\left(  2n\right)  },\label{t3q}%
\end{equation}
where $Q_{\bm{A}\leftarrow\tilde{\bm{A}}\leftarrow\bar{\bm{A}}}^{\left(
2n\right)  }$ is a $2n$-form which depends on all three connections and whose
explicit form cannot be directly determined from the Chern--Weil Theorem
alone. Now we recast the `triangle' equation (\ref{t3q}) in a more suggestive
way as
\begin{equation}
Q_{\bm{A}\leftarrow\bar{\bm{A}}}^{\left(  2n+1\right)  }=Q_{\bm{A}\leftarrow
\tilde{\bm{A}}}^{\left(  2n+1\right)  }+Q_{\tilde{\bm{A}}\leftarrow
\bar{\bm{A}}}^{\left(  2n+1\right)  }+\text{d}Q_{\bm{A}\leftarrow
\tilde{\bm{A}}\leftarrow\bar{\bm{A}}}^{\left(  2n\right)  },\label{tr1}%
\end{equation}
which can be read off as saying that a transgression form `interpolating'
between $\bar{\bm{A}}$ and $\bm{A}$ may be written as the sum of two
transgressions which introduce an intermediate, ancillary one-form
$\tilde{\bm{A}}$ plus a total derivative. It is important to note here that
$\tilde{\bm{A}}$ is completely arbitrary, and may be chosen according to
convenience. Eq.~(\ref{tr1}), used iteratively if necessary, allows us to
split our TGFT lagrangian essentially at wish --- note however that every use
of (\ref{tr1}) brings in a boundary contribution which is so far not known.

In order to obtain an explicit form for $Q_{\bm{A}\leftarrow\tilde
{\bm{A}}\leftarrow\bar{\bm{A}}}^{\left(  2n\right)  }$, and also to show the
common origin of the Chern--Weil Theorem and the Triangle Equation
(\ref{tr1}), we recall here a powerful result known as the Extended Cartan
Homotopy Formula (ECHF)~\cite{Man85}.

\begin{figure}
\includegraphics{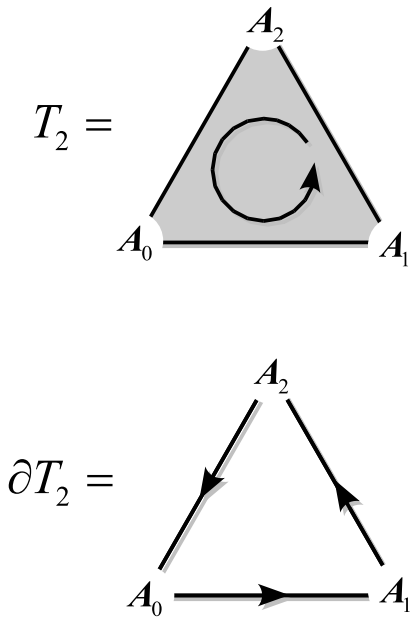}
\caption{\label{mono}Connection structure associated to the simplex vertices.}
\end{figure}

Let us consider a set $\left\{  \bm{A}_{i},i=0,\ldots,r+1\right\}  $ of
one-form gauge connections on a fiber-bundle over a $d$-dimensional manifold
$M$ and a $\left(  r+1\right)  $-dimensional oriented simplex $T_{r+1}$
parameterized by the set $\left\{  t^{i},i=0,\ldots,r+1\right\}  $. These
parameters must satisfy the constraints
\begin{align}
t^{i}  &  \geq0,\qquad i=0,\ldots,r+1,\\
\sum_{i=0}^{r+1}t^{i}  &  =1.\label{tt1}%
\end{align}
Eq.~(\ref{tt1}) in particular implies that the linear combination
\begin{equation}
\bm{A}_{t}=\sum_{i=0}^{r+1}t^{i}\bm{A}_{i}%
\end{equation}
transforms as a gauge connection in the same way as every individual
$\bm{A}_{i}$ does. We can picture each $\bm{A}_{i}$ as associated to the
$i$-th vertex of $T_{r+1}$ (see FIG.~\ref{mono}), which we accordingly denote as
\begin{equation}
T_{r+1}=\left(  \bm{A}_{0}\bm{A}_{1}\cdots\bm{A}_{r+1}\right)  .
\end{equation}

With the preceding notation, the ECHF reads~\cite{Man85}
\begin{align}
\int_{\partial T_{r+1}}\frac{l_{t}^{p}}{p!}\pi &  =\int_{T_{r+1}}\frac
{l_{t}^{p+1}}{\left(  p+1\right)  !}\text{d}\pi+\nonumber\\
&  +\left(  -1\right)  ^{p+q}\text{d}\int_{T_{r+1}}\frac{l_{t}^{p+1}}{\left(
p+1\right)  !}\pi.\label{cehf}%
\end{align}
Here $\pi$ represents a polynomial in the forms $\left\{  \bm{A}_{t}%
,\bm{F}_{t},\text{d}_{t}\bm{A}_{t},\text{d}_{t}\bm{F}_{t}\right\}  $ which is
also an $m$-form on $M$ and a $q$-form on $T_{r+1}$, with $m\geq p$ and
$p+q=r$. The exterior derivatives on $M$ and $T_{r+1}$ are denoted
respectively by d and d$_{t}$. The operator $l_{t}$, called \emph{homotopy
derivation}, maps differential forms on $M$ and $T_{r+1}$ according to
\begin{equation}
l_{t}:\Omega^{a}\left(  M\right)  \times\Omega^{b}\left(  T_{r+1}\right)
\rightarrow\Omega^{a-1}\left(  M\right)  \times\Omega^{b+1}\left(
T_{r+1}\right)  ,
\end{equation}
and it satisfies Leibniz's rule as well as d and d$_{t}$. Its action on
$\bm{A}_{t}$ and $\bm{F}_{t}$ reads~\cite{Man85}%
\begin{align}
l_{t}\bm{F}_{t}  &  =\text{d}_{t}\bm{A}_{t},\label{lfda}\\
l_{t}\bm{A}_{t}  &  =0.
\end{align}

The three operators d, d$_{t}$ and $l_{t}$ define a graded algebra given by
\begin{align}
\text{d}^{2}  &  =0,\label{ga1}\\
\text{d}_{t}^{2}  &  =0,\\
\left[  l_{t},\text{d}\right]   &  =\text{d}_{t},\\
\left[  l_{t},\text{d}_{t}\right]   &  =0,\\
\left\{  \text{d},\text{d}_{t}\right\}   &  =0.\label{ga5}%
\end{align}

Particular cases of (\ref{cehf}), which we review below, reproduce both the
Chern--Weil Theorem, eq.~(\ref{cwt}), and the Triangle Equation, eq.
(\ref{t3q}). In the rest of the paper we will always stick to the polynomial
\begin{equation}
\pi=\left\langle \bm{F}_{t}^{n+1}\right\rangle .
\end{equation}
This choice has the three following properties: (i) $\pi$ is $M$%
-closed~\footnote{This is easily deduced from the invariant property of $\pi$
and Bianchi's identity D$_{t}\bm{F}_{t}=0$.}, i.e., d$\pi=0$, (ii) $\pi$ is a
0-form on $T_{r+1}$, i.e., $q=0$ and (iii) $\pi$ is a $\left(  2n+2\right)
$-form on $M$, i.e., $m=2n+2$. The allowed values for $p$ are $p=0,\ldots
,2n+2$. The ECHF reduces in this case to
\begin{equation}
\int_{\partial T_{p+1}}\frac{l_{t}^{p}}{p!}\left\langle \bm{F}_{t}%
^{n+1}\right\rangle =\left(  -1\right)  ^{p}\text{d}\int_{T_{p+1}}\frac
{l_{t}^{p+1}}{\left(  p+1\right)  !}\left\langle \bm{F}_{t}^{n+1}\right\rangle
.\label{cehfc}%
\end{equation}
We call eq.~(\ref{cehfc}) the `restricted' (or `closed') version of the ECHF.

\subsection{\label{cwt1}$p=0$: Chern--Weil Theorem}

In this section we study the case $p=0$ of eq.~(\ref{cehfc}),
\begin{equation}
\int_{\partial T_{1}}\left\langle \bm{F}_{t}^{n+1}\right\rangle =\text{d}%
\int_{T_{1}}l_{t}\left\langle \bm{F}_{t}^{n+1}\right\rangle ,\label{p0}%
\end{equation}
where we must remember that $\bm{F}_{t}$ is the curvature tensor for the
connection
\begin{equation}
\bm{A}_{t}=t^{0}\bm{A}_{0}+t^{1}\bm{A}_{1},
\end{equation}
with $t^{0}$ and $t^{1}$ satisfying the constraint [cf. eq.~(\ref{tt1})]%
\begin{equation}
t^{0}+t^{1}=1.
\end{equation}
The boundary of the simplex $T_{1}=\left(  \bm{A}_{0}\bm{A}_{1}\right)  $ is
just
\begin{equation}
\partial\left(  \bm{A}_{0}\bm{A}_{1}\right)  =\left(  \bm{A}_{1}\right)
-\left(  \bm{A}_{0}\right)  ,
\end{equation}
so that integration of the {\small LHS} of (\ref{p0}) is trivial:
\begin{equation}
\int_{\partial T_{1}}\left\langle \bm{F}_{t}^{n+1}\right\rangle =\left\langle
\bm{F}_{1}^{n+1}\right\rangle -\left\langle \bm{F}_{0}^{n+1}\right\rangle .
\end{equation}

On the other hand, the symmetric nature of $\left\langle \bm{F}_{t}%
^{n+1}\right\rangle $ implies that
\begin{equation}
l_{t}\left\langle \bm{F}_{t}^{n+1}\right\rangle =\left(  n+1\right)
\left\langle l_{t}\bm{F}_{t}\bm{F}_{t}^{n}\right\rangle .
\end{equation}
Replacing $t^{0}=1-t^{1}$ we may write [cf. eq.~(\ref{lfda})]
\begin{equation}
l_{t}\bm{F}_{t}=\text{d}_{t}\bm{A}_{t}=\text{d}t^{1}\left(  \bm{A}_{1}%
-\bm{A}_{0}\right)  .
\end{equation}
Since integration on $T_{1}$ actually corresponds to integrating with $t^{1}$
from $t^{1}=0$ to $t^{1}=1$, eq.~(\ref{p0}) finally becomes
\begin{equation}
\left\langle \bm{F}_{1}^{n+1}\right\rangle -\left\langle \bm{F}_{0}%
^{n+1}\right\rangle =\text{d}Q_{\bm{A}_{1}\leftarrow\bm{A}_{0}}^{\left(
2n+1\right)  },
\end{equation}
where the transgression form $Q_{\bm{A}_{1}\leftarrow\bm{A}_{0}}^{\left(
2n+1\right)  }$ is defined as
\begin{align}
Q_{\bm{A}_{1}\leftarrow\bm{A}_{0}}^{\left(  2n+1\right)  }  &  \equiv
\int_{T_{1}}l_{t}\left\langle \bm{F}_{t}^{n+1}\right\rangle \nonumber\\
&  =\left(  n+1\right)  \int_{0}^{1}\text{d}t\left\langle \left(
\bm{A}_{1}-\bm{A}_{0}\right)  \bm{F}_{t}^{n}\right\rangle .
\end{align}
This concludes our derivation of the Chern--Weil Theorem as a corollary of the ECHF.

\subsection{$p=1$: Triangle Equation}

In this section we study the case $p=1$ of eq.~(\ref{cehfc}),
\begin{equation}
\int_{\partial T_{2}}l_{t}\left\langle \bm{F}_{t}^{n+1}\right\rangle
=-\text{d}\int_{T_{2}}\frac{l_{t}^{2}}{2}\left\langle \bm{F}_{t}%
^{n+1}\right\rangle ,\label{p1}%
\end{equation}
where $\bm{F}_{t}$ is the curvature corresponding to the connection
$\bm{A}_{t}=t^{0}\bm{A}_{0}+t^{1}\bm{A}_{1}+t^{2}\bm{A}_{2}$. The boundary of
the simplex $T_{2}=\left(  \bm{A}_{0}\bm{A}_{1}\bm{A}_{2}\right)  $ may be
written as the sum
\begin{equation}
\partial\left(  \bm{A}_{0}\bm{A}_{1}\bm{A}_{2}\right)  =\left(  \bm{A}_{1}%
\bm{A}_{2}\right)  -\left(  \bm{A}_{0}\bm{A}_{2}\right)  +\left(
\bm{A}_{0}\bm{A}_{1}\right)  ,
\end{equation}
so that the integral in the {\small LHS} of (\ref{p1}) is decomposed as
\begin{align*}
\int_{\partial T_{2}}l_{t}\left\langle \bm{F}_{t}^{n+1}\right\rangle  &
=\int_{\left(  \bm{A}_{1}\bm{A}_{2}\right)  }l_{t}\left\langle \bm{F}_{t}%
^{n+1}\right\rangle +\\
&  -\int_{\left(  \bm{A}_{0}\bm{A}_{2}\right)  }l_{t}\left\langle
\bm{F}_{t}^{n+1}\right\rangle +\int_{\left(  \bm{A}_{0}\bm{A}_{1}\right)
}l_{t}\left\langle \bm{F}_{t}^{n+1}\right\rangle .
\end{align*}
Each of the terms in this equation is what we called before a transgression
form:
\begin{equation}
\int_{\partial T_{2}}l_{t}\left\langle \bm{F}_{t}^{n+1}\right\rangle
=Q_{\bm{A}_{2}\leftarrow\bm{A}_{1}}^{\left(  2n+1\right)  }-Q_{\bm{A}_{2}%
\leftarrow\bm{A}_{0}}^{\left(  2n+1\right)  }+Q_{\bm{A}_{1}\leftarrow
\bm{A}_{0}}^{\left(  2n+1\right)  }.
\end{equation}

On the other hand, Leibniz's rule for $l_{t}$ and eq.~(\ref{lfda}) imply that
\begin{equation}
\int_{T_{2}}\frac{l_{t}^{2}}{2}\left\langle \bm{F}_{t}^{n+1}\right\rangle
=\frac{1}{2}n\left(  n+1\right)  \int_{T_{2}}\left\langle \left(  \text{d}%
_{t}\bm{A}_{t}\right)  ^{2}\bm{F}_{t}^{n-1}\right\rangle .
\end{equation}
Integrating over the simplex we get
\begin{equation}
\int_{T_{2}}\frac{l_{t}^{2}}{2}\left\langle \bm{F}_{t}^{n+1}\right\rangle
=Q_{\bm{A}_{2}\leftarrow\bm{A}_{1}\leftarrow\bm{A}_{0}}^{\left(  2n\right)  },
\end{equation}
where $Q_{\bm{A}_{2}\leftarrow\bm{A}_{1}\leftarrow\bm{A}_{0}}^{\left(
2n\right)  }$ is given by
\begin{align}
Q_{\bm{A}_{2}\leftarrow\bm{A}_{1}\leftarrow\bm{A}_{0}}^{\left(  2n\right)  }
&  \equiv n\left(  n+1\right)  \int_{0}^{1}dt\int_{0}^{t}ds\nonumber\\
&  \left\langle \left(  \bm{A}_{2}-\bm{A}_{1}\right)  \left(  \bm{A}_{1}%
-\bm{A}_{0}\right)  \bm{F}_{t}^{n-1}\right\rangle .\label{q3}%
\end{align}
In (\ref{q3}) we have introduced dummy parameters $t=1-t^{0}$ and $s=t^{2}$,
in terms of which $\bm{A}_{t}$ reads
\begin{equation}
\bm{A}_{t}=\bm{A}_{0}+t\left(  \bm{A}_{1}-\bm{A}_{0}\right)  +s\left(
\bm{A}_{2}-\bm{A}_{1}\right)  .
\end{equation}

Putting everything together, we find the Triangle Equation
\begin{equation}
Q_{\bm{A}_{2}\leftarrow\bm{A}_{1}}^{\left(  2n+1\right)  }-Q_{\bm{A}_{2}%
\leftarrow\bm{A}_{0}}^{\left(  2n+1\right)  }+Q_{\bm{A}_{1}\leftarrow
\bm{A}_{0}}^{\left(  2n+1\right)  }=-\text{d}Q_{\bm{A}_{2}\leftarrow
\bm{A}_{1}\leftarrow\bm{A}_{0}}^{\left(  2n\right)  },
\end{equation}
or alternatively
\begin{equation}
Q_{\bm{A}_{2}\leftarrow\bm{A}_{0}}^{\left(  2n+1\right)  }=Q_{\bm{A}_{2}%
\leftarrow\bm{A}_{1}}^{\left(  2n+1\right)  }+Q_{\bm{A}_{1}\leftarrow
\bm{A}_{0}}^{\left(  2n+1\right)  }+\text{d}Q_{\bm{A}_{2}\leftarrow
\bm{A}_{1}\leftarrow\bm{A}_{0}}^{\left(  2n\right)  }.\label{treq}%
\end{equation}
We would like to stress here that use of the ECHF has now allowed us to
pinpoint the exact form of the boundary contribution $Q_{\bm{A}_{2}%
\leftarrow\bm{A}_{1}\leftarrow\bm{A}_{0}}^{\left(  2n\right)  }$,
eq.~(\ref{q3}).

\section{\label{trac}Subspace Separation Method for TGFT, with Two Examples}

Our separation method is based on the Triangle Equation (\ref{treq}),
and embodies the following steps:
\begin{enumerate}
\item Identify the relevant subspaces present in the gauge algebra, i.e.,
write $\mathfrak{g}=V_{0}\oplus\cdots\oplus V_{p}$.

\item Write the connections as a sum of pieces valued on every subspace, i.e.,
$\bm{A}=\bm{a}_{0}+\cdots+\bm{a}_{p}$,
$\bar{\bm{A}}=\bar{\bm{a}}_{0}+\cdots+\bar{\bm{a}}%
_{p}$.

\item Use eq. (\ref{treq}) with
\begin{align}
\bm{A}_{0}  & =\bar{\bm{A}},\\
\bm{A}_{1}  & =\bm{a}_{0}+\cdots+\bm{a}_{p-1},\\
\bm{A}_{2}  & =\bm{A}.
\end{align}

\item Repeat step 3 for the transgression $Q_{\bm{A}_{1}%
\leftarrow\bm{A}_{0}}$, etc.
\end{enumerate}

After performing these steps, one ends up with an equivalent expression
for the TGFT lagrangian which has been separated in two different ways.
First, the lagrangian is split into bulk and boundary contributions. This
is due to the fact that each use of eq. (\ref{treq}) brings in a new
boundary term. Second, each term in the bulk lagrangian refers to
a different subspace of the gauge algebra. This comes about because
the difference $\bm{A}_{2}-\bm{A}_{1}$ is valued only on one particular
subspace.

Below we show two examples of TGFTs, one for Gravity and one for SUGRA. In
both cases the separation method is used to cast the lagrangian in a
physically sensible, readable way.

\subsection{\label{fapg}Finite Action Principle for Gravity}

In this section we aim to show explicitly how the lagrangian for gravity in
$d=2n+1$ given in~\cite{Mor04} corresponds to a transgression form for the
connections
\begin{align}
\bm{A}_{0}  &  =\bar{\bm{\omega}},\label{Abar}\\
\bm{A}_{2}  &  =\bm{e}+\bm{\omega}.\label{A}%
\end{align}
Here we use the abbreviations $\bm{e}=e^{a}\bm{P}_{a}$, $\bm{\omega}=\frac
{1}{2}\omega^{ab}\bm{J}_{ab}$, $\bar{\bm{\omega}}=\frac{1}{2}\bar{\omega}%
^{ab}\bm{J}_{ab}$, with $\bm{P}_{a},\bm{J}_{ab}$ being the generators of the
AdS Algebra $\mathfrak{so}\left(  2n,2\right)  $. The curvatures for these
connections read
\begin{align}
\bm{F}_{0}  &  =\bar{\bm{R}},\\
\bm{F}_{2}  &  =\bm{R}+\bm{e}^{2}+\bm{T},
\end{align}
where
\begin{align}
\bm{R}  &  =\text{d}\bm{\omega}+\bm{\omega}^{2},\label{R}\\
\bm{T}  &  =\text{d}\bm{e}+\left[  \bm{\omega },\bm{e}\right]  ,
\end{align}
are the Lorentz curvature and the torsion, respectively [an expression
completely analogous to (\ref{R}) is valid for $\bar{\bm{R}}$].

As an invariant polynomial for the AdS Algebra we shall stick to our previous
choice of the Levi-Civita tensor [cf. eq.~(\ref{eps})],
\begin{equation}
\left\langle \bm{J}_{a_{1}a_{2}}\cdots\bm{J}_{a_{2n-1}a_{2n}}\bm{P}_{a_{2n+1}%
}\right\rangle =\frac{2^{n}}{\ell}\varepsilon_{a_{1}\cdots a_{2n+1}%
},\label{eps2}%
\end{equation}
with all other possible combinations vanishing.

In order to separate the pieces of our TGFT lagrangian in a meaningful way, we
introduce the intermediate connection
\begin{equation}
\bm{A}_{1}=\bm{\omega}
\end{equation}
and consider the Triangle Equation (\ref{treq}) as follows:
\begin{equation}
L_{\bm{e}+\bm{\omega}\leftarrow\bar{\bm{\omega}}}^{\left(  2n+1\right)
}=L_{\bm{e}+\bm{\omega}\leftarrow\bm{\omega}}^{\left(  2n+1\right)
}+kQ_{\bm{\omega}\leftarrow\bar{\bm{\omega}}}^{\left(  2n+1\right)
}+k\text{d}Q_{\bm{e}+\bm{\omega}\leftarrow\bm{\omega}\leftarrow\bar
{\bm{\omega}}}^{\left(  2n\right)  }.\label{trb}%
\end{equation}
From section~\ref{llg} we know that the first term in the {\small RHS} of
(\ref{trb}) corresponds to a LL lagrangian with the canonical coefficients
[cf. eq.~(\ref{lcst})],
\begin{equation}
L_{\bm{e}+\bm{\omega}\leftarrow\bm{\omega}}^{\left(  2n+1\right)  }=\left(
n+1\right)  k\int_{0}^{1}dt\left\langle \bm{eF}_{t}^{n}\right\rangle ,
\end{equation}
with $\bm{F}_{t}$ being the curvature for the connection $\bm{A}_{t}%
=\bm{\omega}+t\bm{e}$,$\ \bm{F}_{t}=\bm{R}+t\bm{T}+t^{2}\bm{e}^{2}$.

Our particular choice for the invariant polynomial now implies that the second
term in (\ref{trb}) vanishes:
\begin{equation}
Q_{\bm{\omega}\leftarrow\bar{\bm{\omega}}}^{\left(  2n+1\right)  }=0.
\end{equation}

Going back to eq.~(\ref{q3}) we find that the boundary contribution in
(\ref{trb}) may be written as
\begin{equation}
Q_{\bm{e}+\bm{\omega}\leftarrow\bm{\omega}\leftarrow\bar{\bm{\omega}}%
}^{\left(  2n\right)  }=n\left(  n+1\right)  \int_{0}^{1}dt\int_{0}%
^{t}ds\left\langle \bm{e\theta F}_{st}^{n-1}\right\rangle ,
\end{equation}
where
\begin{equation}
\bm{\theta}\equiv\bm{\omega}-\bar{\bm{\omega}}%
\end{equation}
and$\ \bm{F}_{st}$ is the curvature~\footnote{The easiest way to compute this
curvature makes use of (a generalized version of) the Gauss--Codazzi
equations. Let $\bm{A}$ and $\bar{\bm{A}}$ be two one-form gauge connections
and let $\bm{\Delta}=\bm{A}-\bar{\bm{A}}$. Then the corresponding curvatures
are related by $\bm{F}=\bar{\bm{F}}+\text{\={D}}\bm{\Delta }+\bm{\Delta}^{2}$,
where \={D} is the covariant derivative in the connection $\bar{\bm{A}}$.} for
the connection $\bm{A}_{st}=\bar{\bm{\omega}}+s\bm{e}+t\bm{\theta}$,
\begin{equation}
\bm{F}_{st}=\bar{\bm{R}}+\text{D}_{\bar{\bm{\omega}}}\left(
s\bm{e}+t\bm{\theta}\right)  +s^{2}\bm{e}^{2}+st\left[
\bm{e},\bm{\theta}\right]  +t^{2}\bm{\theta}^{2}.
\end{equation}

Putting everything together, our final lagrangian reads
\begin{align}
L_{\bm{e}+\bm{\omega}\leftarrow\bar{\bm{\omega}}}^{\left(  2n+1\right)  }  &
=\left(  n+1\right)  k\int_{0}^{1}dt\left\langle \bm{e}\left(  \bm{R}+t^{2}%
\bm{e}^{2}\right)  ^{n}\right\rangle +\nonumber\\
&  +k\text{d}Q_{\bm{e}+\bm{\omega}\leftarrow\bm{\omega}\leftarrow
\bar{\bm{\omega}}}^{\left(  2n\right)  }.\label{fgl}%
\end{align}

The field equations for (\ref{fgl}) are given by
\begin{align}
\left\langle \bm{J}_{ab}\left(  \bm{R}+\bm{e}^{2}\right)  ^{n-1}%
\bm{T}\right\rangle  &  =0,\label{jrt}\\
\left\langle \bm{P}_{a}\left(  \bm{R}+\bm{e}^{2}\right)  ^{n}\right\rangle  &
=0.\label{pre}%
\end{align}
These can be obtained by direct variation of (\ref{fgl}) or by replacing
$\bm{F}_{0}$ and $\bm{F}_{2}$ in (\ref{fg1})--(\ref{fg2}). A more explicit
version is found making use of (\ref{eps2}):
\begin{align}
0  &  =\varepsilon_{aba_{1}\cdots a_{2n-1}}\left(  R^{a_{1}a_{2}}+\frac
{1}{\ell^{2}}e^{a_{1}}e^{a_{2}}\right)  \times\cdots\times\nonumber\\
&  \times\left(  R^{a_{2n-3}a_{2n-2}}+\frac{1}{\ell^{2}}e^{a_{2n-3}%
}e^{a_{2n-2}}\right)  T^{a_{2n-1}},
\end{align}%
\begin{align}
0  &  =\varepsilon_{aa_{1}\cdots a_{2n}}\left(  R^{a_{1}a_{2}}+\frac{1}%
{\ell^{2}}e^{a_{1}}e^{a_{2}}\right)  \times\nonumber\\
&  \times\cdots\times\left(  R^{a_{2n-1}a_{2n}}+\frac{1}{\ell^{2}}e^{a_{2n-1}%
}e^{a_{2n}}\right)  .
\end{align}

We have again two choices for obtaining boundary conditions; by direct
variation of (\ref{fgl}) or by replacing the relevant quantities in our
general formula, eq.~(\ref{bc}). Any of them can be shown to yield
\begin{equation}
\left.  \int_{0}^{1}dt\left\langle \left(  \delta\bar{\bm{\omega}}%
+t\delta\bm{\theta}+t\delta\bm{e}\right)  \left(  \bm{\theta}+\bm{e}\right)
\bm{F}_{t}^{n-1}\right\rangle \right\vert _{\partial M}=0,\label{bct}%
\end{equation}
where in this case the connection $\bm{A}_{t}$ and the corresponding curvature
$\bm{F}_{t}$ are given by
\begin{align}
\bm{A}_{t}  &  =\bar{\bm{\omega}}+t\left(  \bm{e}+\bm{\theta}\right)  ,\\
\bm{F}_{t}  &  =\bar{\bm{R}}+t\text{D}_{\bar{\bm{\omega}}}\left(
\bm{e}+\bm{\theta}\right)  +t^{2}\left(  \bm{e}^{2}+\left[
\bm{e},\bm{\theta }\right]  +\bm{\theta}^{2}\right)  .
\end{align}

There are many alternative ways of satisfying boundary conditions (\ref{bct}).
In~\cite{Mor04}, physical arguments~\footnote{The main physical requirement is
the equivalence of the concept of parallel transport induced by $\omega$ and
$\bar{\omega}$ on the boundary of the space-time manifold $M$.} are given that
allow to partially fix the boundary conditions; perhaps the most significant
of them is demanding that $\bar{\bm{\omega}}$ have a fixed value on $\partial
M$, i.e.,
\begin{equation}
\left.  \delta\bar{\bm{\omega}}\right\vert _{\partial M}=0.
\end{equation}
The remaining boundary conditions may be written as
\begin{equation}
\left.  \int_{0}^{1}dt\left\langle t\left(  \delta
\bm{\theta e}-\bm{\theta}\delta\bm{e}\right)  \left(  \bar{\bm {R}}%
+t^{2}\bm{e}^{2}+t^{2}\bm{\theta}^{2}\right)  ^{n-1}\right\rangle \right\vert
_{\partial M}=0,
\end{equation}
and can be readily satisfied by requiring
\begin{equation}
\delta\theta^{\lbrack ab}e^{c]}=\theta^{\lbrack ab}\delta e^{c]}.
\end{equation}

We would like to stress that the choice (\ref{eps2})\ for the invariant
polynomial sends all dependence on $\bar{\bm{\omega}}$ in the action to a
boundary term, and in this way the potential conflict of having two
independent CS theories living on the same space-time manifold is avoided. The
presence of $\bar{\bm{\omega}}$ in the lagrangian does nevertheless have a
dramatic effect on the theory, as it changes the boundary conditions and
renders both the TGFT action and the conserved charges finite~\footnote{As
shown in~\cite{Mor04,Ole04}, the same boundary term may be used to render the
EH action in higher odd dimensions finite.}. Further important theoretical
implications are examined in section~\ref{sur}.

\subsection{Supergravity as a TGFT}

The move from standard CS Theory to TGFT earns us several important
advantages, both in computational power and in theoretical clarity. The latter
will be thoroughly discussed in section~\ref{sur}; here we shall be mainly
concerned with elaborating on the former.

One first disadvantage of the standard CS action formula surfaces when one
wants to perform the separation of the lagrangian in reflection of the
subspaces structure of the gauge algebra. As a matter of fact, it is clear
from the very nature of a CS form that this will require intensive use of
Leibniz's rule, which, especially for complicated algebras in dimensions
higher than three, renders the task a highly non-trivial `artistic' work.

These manipulations finally lead to a separation of the CS lagrangian in bulk and
boundary contributions. After performing this separation, it is no longer
clear whether one should simply drop these boundary terms, since the bulk
lagrangian is still invariant under infinitesimal gauge transformations (up to a total
derivative). Even more involved is the derivation of boundary conditions from
the CS lagrangian; on one hand, they're, from a purely computational point of
view, rather difficult to extract, and on the other, they of course depend on
our earlier choice of dropping or not the boundary contributions just obtained.

On the other hand, the TGFT lagrangian clearly distinguishes itself from a CS
lagrangian in this respect. As a matter of fact, the separation method sketched
at the beginning of section~\ref{trac} applied to the TGFT lagrangian
permits the straightforward realization of both tasks. The lagrangian is split
into bulk and boundary contributions, and the bulk sector is divided into pieces
that faithfully reflect the subspace structure of the gauge algebra. Furthermore,
the boundary conditions arising from these boundary terms have a chance to be
physically meaningful, due to the full invariance of the TGFT lagrangian under
gauge transformations.

In order to highlight the way in which the TGFT formalism deals with these
issues, we present here the TGFT derivation of a CS SUGRA lagrangian. For a
non-trivial example we pick $d=5$ and choose the $\mathcal{N}$-extended AdS
superalgebra $\mathfrak{u}\left(  4|\mathcal{N}\right)  $ (see Refs.
\cite{Cha90,Tro96,Tro98}). This algebra is generated by the set $\left\{
\bm{K},\bm{P}_{a},\bm{J}_{ab},\bm{M}_{n}^{\;m},\bm{Q}_{i}^{\alpha}%
,\bar{\bm{Q}}_{\alpha}^{i}\right\}  $. Latin letters from the beginning of the
alphabet $\left(  a,b,c,\ldots\right)  $ denote Lorentz indices and rank from
0 to 4; Greek letters $\left(  \alpha,\beta,\gamma,\ldots\right)  $ denote
spinor indices and rank from 1 to 4 (the dimension of a Dirac spinor in $d=5$
is $2^{\left[  5/2\right]  }=4$); Latin letters from the middle of the
alphabet $\left(  i,j,k,\ldots\right)  $ denote $\mathfrak{su}\left(
\mathcal{N}\right)  $ indices, which rank from 1 to $\mathcal{N}$ and can be
regarded as `counting' fermionic generators. The non-vanishing
(anti)commutation relations read
\begin{align}
\left[  \bm{K},\bm{Q}_{k}^{\alpha}\right]   &  =-i\left(  \frac{1}{4}-\frac
{1}{\mathcal{N}}\right)  \bm{Q}_{k}^{\alpha},\\
\left[  \bm{M}_{k}^{\;j},\bm{Q}_{m}^{\alpha}\right]   &  =\left(  \delta
_{k}^{p}\delta_{m}^{j}-\frac{1}{\mathcal{N}}\delta_{m}^{p}\delta_{k}%
^{j}\right)  \bm{Q}_{p}^{\alpha},
\end{align}
\begin{align}
\left[  \bm{P}_{a},\bm{Q}_{k}^{\alpha}\right]   &  =-\frac{1}{2\ell}\left(
\Gamma_{a}\right)  _{\;\beta}^{\alpha}\bm{Q}_{k}^{\beta},\\
\left[  \bm{J}_{ab},\bm{Q}_{k}^{\alpha}\right]   &  =-\frac{1}{2}\left(
\Gamma_{ab}\right)  _{\;\beta}^{\alpha}\bm{Q}_{k}^{\beta},
\end{align}%
\begin{equation}
\left[  \bm{M}_{k}^{\;j},\bm{M}_{m}^{\;l}\right]  =\delta_{m}^{j}%
\bm{M}_{k}^{\;l}-\delta_{k}^{l}\bm{M}_{m}^{\;j},
\end{equation}%
\begin{align}
\left\{  \bm{Q}_{k}^{\alpha},\bar{\bm{Q}}_{\beta}^{j}\right\}   &
=2\delta_{k}^{j}\left(  \Gamma^{a}\right)  _{\;\beta}^{\alpha}\bm{P}_{a}%
+\nonumber\\
&  -\frac{4}{\ell}\left[  i\delta_{k}^{j}\delta_{\beta}^{\alpha}%
\bm{K}-\delta_{\beta}^{\alpha}\bm{M}_{k}^{\;j}+\frac{1}{4}\delta_{k}%
^{j}\left(  \Gamma^{ab}\right)  _{\;\beta}^{\alpha}\bm{J}_{ab}\right]
,\label{qq}%
\end{align}
where $\Gamma_{a}$ are Dirac Matrices~\footnote{There are two inequivalent
representations for the Dirac Matrices in $d=5$. In order to distinguish
between them we define $\gamma\equiv\left(  1/4i\right)  \operatorname*{Tr}%
\left(  \Gamma\right)  =\pm1$, where $\Gamma\equiv\Gamma_{0}\cdots\Gamma_{4}%
$.} in $d=5$ and $\Gamma_{ab}\equiv\frac{1}{2}\left[  \Gamma_{a},\Gamma
_{b}\right]  $. The generators $\bm{M}_{k}^{\;j}$ span an $\mathfrak{su}%
\left(  \mathcal{N}\right)  $ subalgebra, while $\bm{P}_{a}$ and $\bm{J}_{ab}$
generate as usual the AdS algebra [omitted above, see (\ref{PP})--(\ref{JJ0}%
)]. The anticommutator $\left\{  \bm{Q},\bar{\bm{Q}}\right\}  $ has components
on all bosonic generators, and not only on the translational ones. This is a
consequence of the fact that we are considering the supersymmetric version of
the AdS Algebra, as opposed to the Poincar\'{e} Algebra. The latter may be
recovered via an \.{I}n\"{o}n\"{u}--Wigner contraction by setting
$\ell\rightarrow\infty$ [in which case only the translational part in the
{\small RHS} of (\ref{qq}) survives]. Commutators of the form $\left[
\bm{B},\bar{\bm{Q}}\right]  $, where $\bm{B}$ is some bosonic generator,
differ only by a sign from their $\left[  \bm{B},\bm{Q}\right]  $ counterparts
and have not been explicitly written out. For $\mathcal{N}=4$ the generator
$\bm{K}$ becomes abelian and factors out from the rest of the algebra.

The second ingredient we need in order to write down the TGFT lagrangian is a
$\mathfrak{u}\left(  4|\mathcal{N}\right)  $-invariant symmetric polynomial
$\left\langle \cdots\right\rangle $ of rank three. This can be conveniently
defined as the supersymmetrized supertrace of the product of three
supermatrices representing as many generators in $\mathfrak{u}\left(
4|\mathcal{N}\right)  $. The fact that the Dirac Matrices provide a natural
representation for the AdS Algebra is a turning point that makes this
construction feasible. Without going into the details, the invariant
polynomial we will use satisfies the following contraction identities:
\begin{eqnarray}
\left\langle \bm{K}^{3} \right\rangle & = & \frac{1}{4^{2}} - 
\frac{1}{\mathcal{N}^{2}},\\
B_{\;n}^{m} C_{\;q}^{p} \left\langle \bm{KM}_{m}^{\;n} \bm{M}_{p}^{\;q}
\right\rangle & = & \frac{1}{\mathcal{N}} B_{\;p}^{q} C_{\;q}^{p},\\
B^{a}C^{b} \left\langle \bm{KP}_{a} \bm{P}_{b} \right\rangle & = & - 
\frac{1}{4\ell^{2}} B^{a}C_{a},\\
B^{ab} C^{cd} \left\langle \bm{KJ}_{ab} \bm{J}_{cd} \right\rangle & = & - 
\frac{1}{2} B_{\;b}^{a} C_{\;a}^{b},
\end{eqnarray}
\begin{align}
A_{\;n}^{m} B_{\;q}^{p} B_{\;s}^{r} \left\langle \bm{M}_{m}^{\;n}
\bm{M}_{p}^{\;q} \bm{M}_{r}^{\;s} \right\rangle & = -i A_{\;p}^{m}
B_{\;q}^{p}B_{\;m}^{q},\\
A^{a} B^{bc} C^{de} \left\langle \bm{P}_{a} \bm{J}_{bc} \bm{J}_{de} \right\rangle
& = -\frac{\gamma}{2\ell}\varepsilon_{abcde}A^{a}B^{bc}C^{de},\label{eps3}
\end{align}
\begin{eqnarray}
\bar{\zeta}^{m} B \chi_{n} \left\langle \bm{Q}_{m} \bm{K}\bar{\bm{Q}}^{n} \right\rangle
& = & \frac{2}{\ell} \left( \frac{1}{4} + \frac{1}{\mathcal{N}}\right) 
\bar{\zeta} B \chi,\\
\bar{\zeta}^{m} B^{a} \chi_{n} \left\langle \bm{Q}_{m} \bm{J}_{a}
\bar{\bm{Q}}^{n} \right\rangle & = & -\frac{i}{\ell} \bar{\zeta} B^{a} \Gamma_{a}\chi,\\
\bar{\zeta}^{m} B^{ab} \chi_{n} \left\langle \bm{Q}_{m} \bm{J}_{ab}
\bar{\bm{Q}}^{n} \right\rangle & = & -\frac{i}{\ell} \bar{\zeta} B^{ab}
\Gamma_{ab}\chi,\\
\bar{\zeta}^{p} B_{\;n}^{m} \chi_{q} \left\langle \bm{Q}_{p} \bm{M}_{m}^{\;n}
\bar{\bm{Q}}^{q} \right\rangle & = & -\frac{2i}{\ell} \bar{\zeta}^{n}
B_{\;n}^{m}\chi_{m}.
\end{eqnarray}
Here $A,B,C,\zeta,\chi$ are arbitrary differential forms with appropriate
index structure~\footnote{Following standard practice, we usually omit spinor
indices (especially when summed). Typical shortcuts are $\bm{\psi}_{j}%
=\psi_{j\mu}^{\alpha}\bar{\bm{Q}}_{\alpha}$d$x^{\mu}$, $\bar{\chi}\psi
=\bar{\chi}_{\alpha}^{k}\psi_{k}^{\alpha}$, $\bar{\chi}\Gamma_{a}\psi
=\bar{\chi}_{\alpha}^{k}\left(  \Gamma_{a}\right)  _{\;\beta}^{\alpha}\psi
_{k}^{\beta}$, etc.}. It is interesting to note that the invariant polynomial
used in section~\ref{fapg} is recovered unchanged (but for a different overall
factor) in eq.~(\ref{eps3}).

We will choose as lagrangian the transgression form that interpolates between
the following connections:
\begin{align}
\bm{A}_{0}  &  =\bar{\bm{\omega}},\\
\bm{A}_{4}  &  =\bm{e}+\bm{\omega}+\bm{b}+\bm{a}+\bar{\bm{\psi}}-\bm{\psi},
\end{align}
where
\begin{align}
\bm{e}  &  =e^{a}\bm{P}_{a},\\
\bm{\omega}  &  =\frac{1}{2}\omega^{ab}\bm{J}_{ab},\\
\bar{\bm{\omega}}  &  =\frac{1}{2}\bar{\omega}^{ab}\bm{J}_{ab},\\
\bm{b}  &  =b\bm{K},\\
\bm{a}  &  =a_{\;n}^{m}\bm{M}_{m}^{\;n},\\
\bar{\bm{\psi}}  &  =\bar{\psi}\bm{Q},\\
\bm{\psi}  &  =\bar{\bm{Q}}\psi.
\end{align}
The corresponding curvatures read
\begin{equation}
\bm{F}_{0}=\bar{\bm{R}},
\end{equation}%
\begin{equation}
\bm{F}_{4}=\left(  \text{d}b+\frac{i}{\ell}\bar{\psi}\psi\right)
\bm{K}+\bm{f}^{\prime}+\bm{T}^{\prime}+\bm{R}^{\prime}+\frac{1}{2}\left(
\nabla\bar{\psi}\bm{Q}-\bar{\bm {Q}}\nabla\psi\right)  ,
\end{equation}
where we have defined
\begin{align}
\bm{f}  &  =\text{d}\bm{a}+\bm{a}^{2},\\
\bm{f}^{\prime}  &  =\bm{f}-\frac{1}{\ell}\left(  \bar{\psi}^{m}\psi
_{n}\right)  \bm{M}_{m}^{\;n},\\
\bm{T}^{\prime}  &  =\bm{T}-\frac{1}{2}\left(  \bar{\psi}\Gamma^{a}%
\psi\right)  \bm{P}_{a},\\
\bm{R}^{\prime}  &  =\bm{R}+\bm{e}^{2}+\frac{1}{4\ell}\left(  \bar{\psi}%
\Gamma^{ab}\psi\right)  \bm{J}_{ab},
\end{align}
and
\begin{align}
\nabla\psi_{k}  &  \equiv\text{d}\psi_{k}+\frac{1}{2\ell}e^{a}\Gamma_{a}%
\psi_{k}+\frac{1}{4}\omega^{ab}\Gamma_{ab}\psi_{k}+\nonumber\\
&  -a_{\;k}^{l}\psi_{l}+i\left(  \frac{1}{4}-\frac{1}{\mathcal{N}}\right)
b\psi_{k},
\end{align}%
\begin{align}
\nabla\bar{\psi}^{k}  &  \equiv\text{d}\bar{\psi}^{k}-\frac{1}{2\ell}e^{a}%
\bar{\psi}^{k}\Gamma_{a}-\frac{1}{4}\omega^{ab}\bar{\psi}^{k}\Gamma
_{ab}+\nonumber\\
&  +a_{\;l}^{k}\bar{\psi}^{l}-i\left(  \frac{1}{4}-\frac{1}{\mathcal{N}%
}\right)  b\bar{\psi}^{k}.
\end{align}

We thus have
\begin{align}
L_{\text{sugra}}^{\left(  5\right)  }  &  =kQ_{\bm{A}_{4}\leftarrow\bm{A}_{0}%
}^{\left(  5\right)  },\nonumber\\
&  =3k\int_{0}^{1}dt\left\langle \bm{\Theta F}_{t}^{2}\right\rangle
,\label{lsg5}%
\end{align}
with
\begin{align}
\bm{\theta}  &  =\bm{\omega}-\bar{\bm{\omega}},\\
\bm{\Theta}  &  =\bm{e}+\bm{\theta}+\bm{b}+\bm{a}+\bar{\bm{\psi}}-\bm{\psi},
\end{align}
and $\bm{F}_{t}$ being the curvature for the connection
\begin{equation}
\bm{A}_{t}=\bar{\bm{\omega}}+t\left(  \bm{e}+\bm{\theta}+\bm{b}+\bm{a}+\bar
{\bm{\psi}}-\bm{\psi}\right)  .
\end{equation}

In order to make sense of (\ref{lsg5}) we introduce the following set of
intermediate connections:
\begin{align}
\bm{A}_{1}  &  =\bm{e}+\bm{\omega},\\
\bm{A}_{2}  &  =\bm{e}+\bm{\omega}+\bm{b},\\
\bm{A}_{3}  &  =\bm{e}+\bm{\omega}+\bm{b}+\bm{a}.
\end{align}
Now we split $Q_{\bm{A}_{4}\leftarrow\bm{A}_{0}}^{\left(  5\right)  }$
according to the pattern~\footnote{Eqs.~(\ref{pat1})--(\ref{pat3}) show of
course only one among many possible splittings.}
\begin{align}
Q_{\bm{A}_{4}\leftarrow\bm{A}_{0}}^{\left(  5\right)  }  &  =Q_{\bm{A}_{4}%
\leftarrow\bm{A}_{3}}^{\left(  5\right)  }+Q_{\bm{A}_{3}\leftarrow\bm{A}_{0}%
}^{\left(  5\right)  }+\text{d}Q_{\bm{A}_{4}\leftarrow\bm{A}_{3}%
\leftarrow\bm{A}_{0}}^{\left(  4\right)  },\label{pat1}\\
Q_{\bm{A}_{3}\leftarrow\bm{A}_{0}}^{\left(  5\right)  }  &  =Q_{\bm{A}_{3}%
\leftarrow\bm{A}_{2}}^{\left(  5\right)  }+Q_{\bm{A}_{2}\leftarrow\bm{A}_{0}%
}^{\left(  5\right)  }+\text{d}Q_{\bm{A}_{3}\leftarrow\bm{A}_{2}%
\leftarrow\bm{A}_{0}}^{\left(  4\right)  },\\
Q_{\bm{A}_{2}\leftarrow\bm{A}_{0}}^{\left(  5\right)  }  &  =Q_{\bm{A}_{2}%
\leftarrow\bm{A}_{1}}^{\left(  5\right)  }+Q_{\bm{A}_{1}\leftarrow\bm{A}_{0}%
}^{\left(  5\right)  }+\text{d}Q_{\bm{A}_{2}\leftarrow\bm{A}_{1}%
\leftarrow\bm{A}_{0}}^{\left(  4\right)  }.\label{pat3}%
\end{align}
Picking up the pieces, we are left with
\begin{equation}
L_{\text{sugra}}^{\left(  5\right)  }=L_{\bm{\psi}}^{\left(  5\right)
}+L_{\bm{a}}^{\left(  5\right)  }+L_{\bm{b}}^{\left(  5\right)  }%
+L_{\bm{e}+\bm{\omega}\leftarrow\bar{\bm{\omega}}}^{\left(  5\right)
}+\text{d}B_{\text{sugra}}^{\left(  4\right)  },\label{lsg5p}%
\end{equation}
where we have defined
\begin{align}
L_{\bm{\psi}}^{\left(  5\right)  }  &  \equiv kQ_{\bm{A}_{4}\leftarrow
\bm{A}_{3}}^{\left(  5\right)  },\\
L_{\bm{a}}^{\left(  5\right)  }  &  \equiv kQ_{\bm{A}_{3}\leftarrow\bm{A}_{2}%
}^{\left(  5\right)  },\\
L_{\bm{b}}^{\left(  5\right)  }  &  \equiv kQ_{\bm{A}_{2}\leftarrow\bm{A}_{1}%
}^{\left(  5\right)  },\\
L_{\bm{e}+\bm{\omega}\leftarrow\bar{\bm{\omega}}}^{\left(  5\right)  }  &
\equiv kQ_{\bm{A}_{1}\leftarrow\bm{A}_{0}}^{\left(  5\right)  },
\end{align}%
\begin{equation}
B_{\text{sugra}}^{\left(  4\right)  }\equiv kQ_{\bm{A}_{2}\leftarrow
\bm{A}_{1}\leftarrow\bm{A}_{0}}^{\left(  4\right)  }+kQ_{\bm{A}_{3}%
\leftarrow\bm{A}_{2}\leftarrow\bm{A}_{0}}^{\left(  4\right)  }+kQ_{\bm{A}_{4}%
\leftarrow\bm{A}_{3}\leftarrow\bm{A}_{0}}^{\left(  4\right)  }.
\end{equation}
A few comments are in order. Ignoring for the moment the boundary contribution
$B_{\text{sugra}}^{\left(  4\right)  }$, we see that all dependence on the
fermions has been packaged in $L_{\bm{\psi}}^{\left(  5\right)  }$, which we
call `fermionic lagrangian'. Similarly, $L_{\bm{a}}^{\left(  5\right)  }$ and
$L_{\bm{b}}^{\left(  5\right)  }$ correspond to pieces that are highly
dependent on $\bm{a}$ and $\bm{b}$ respectively, although some dependence on
$\bm{a}$ and $\bm{b}$ is also found on $L_{\bm{\psi}}^{\left(  5\right)  }$.
In turn, $L_{\bm{b}}^{\left(  5\right)  }$ carries no dependence on $\bm{a}$.
The last piece, $L_{\bm{e}+\bm{\omega
}\leftarrow\bar{\bm{\omega}}}^{\left(  5\right)  }$, is, thanks to the
particular invariant polynomial used, exactly what we considered in section
\ref{fapg}, and stands alone as the `gravity' lagrangian. Explicit versions
for every piece may be easily obtained by going back to the definition of a
transgression form, eq.~(\ref{tra}). As a matter of fact, a straightforward
computation yields
\begin{align}
L_{\bm{\psi}}^{\left(  5\right)  }  &  =\frac{3k}{2i}\left(  \bar{\psi
}_{\alpha}\mathcal{R}_{\;\beta}^{\alpha}\nabla\psi^{\beta}+\bar{\psi}%
^{n}\mathcal{F}_{\;n}^{m}\nabla\psi_{m}+\right. \nonumber\\
&  \left.  -\nabla\bar{\psi}_{\alpha}\mathcal{R}_{\;\beta}^{\alpha}\psi
^{\beta}-\nabla\bar{\psi}^{n}\mathcal{F}_{\;n}^{m}\psi_{m}\right)  ,
\end{align}%
\begin{align}
L_{\bm{a}}^{\left(  5\right)  }  &  =\frac{3k}{\mathcal{N}}\left(
\text{d}b\right)  \operatorname*{Tr}\left(  a\text{d}a+\frac{2}{3}%
a^{3}\right)  +\nonumber\\
&  -ik\operatorname*{Tr}\left[  a\left(  \text{d}a\right)  ^{2}+\frac{3}%
{2}a^{3}\text{d}a+\frac{3}{5}a^{5}\right]  ,
\end{align}%
\begin{align}
L_{\bm{b}}^{\left(  5\right)  }  &  =k\left(  \frac{1}{4^{2}}-\frac
{1}{\mathcal{N}^{2}}\right)  b\left(  \text{d}b\right)  ^{2}+\nonumber\\
&  -\frac{3k}{4\ell^{2}}b\left(  T^{a}T_{a}-R_{ab}e^{a}e^{b}-\frac{\ell^{2}%
}{2}R^{ab}R_{ab}\right)  ,
\end{align}
where
\begin{align}
\mathcal{R}_{\;\beta}^{\alpha}  &  =i\left(  \frac{1}{4}+\frac{1}{\mathcal{N}%
}\right)  \left(  \text{d}b+\frac{i}{2\ell}\bar{\psi}\psi\right)
\delta_{\beta}^{\alpha}+\nonumber\\
&  +\frac{1}{2}\left(  T^{a}-\frac{1}{4}\bar{\psi}\Gamma^{a}\psi\right)
\left(  \Gamma_{a}\right)  _{\;\beta}^{\alpha}+\nonumber\\
&  +\frac{1}{4}\left(  R^{ab}+\frac{1}{\ell^{2}}e^{a}e^{b}+\frac{1}{4\ell}%
\bar{\psi}\Gamma^{ab}\psi\right)  \left(  \Gamma_{ab}\right)  _{\;\beta
}^{\alpha},
\end{align}%
\begin{equation}
\mathcal{F}_{\;n}^{m}=f_{\;n}^{m}-\frac{1}{2\ell}\bar{\psi}^{m}\psi_{n}.
\end{equation}
The lagrangian for the $\mathfrak{su}\left(  \mathcal{N}\right)  $ field
$\bm{a}$ includes both a CS term for $d=5$ and a CS term for $d=3$, the latter
being suitable multiplied by the field-strength for the $\bm{b}$-field, d$b$.

The equations of motion and the boundary conditions are easily obtained using
the general expressions (\ref{fg1})--(\ref{bc}). These are natural extensions
of (\ref{jrt})--(\ref{pre}) and are also found in~\cite{Tro98}.

We would like to stress the fact that the preceding results have been obtained
in a completely straightforward way, without using Leibniz's rule, following
the method given in section \ref{trac}. The same task can be painstakingly long
if approached na\"{\i}vely, i.e., through the sole use of Leibniz's rule and the
definition of a CS or TGFT lagrangian.

\section{\label{sur}Discussion}

\subsection{The Gauge-Theory Structure of TGFT Gravity}

In the TGFT lagrangians for gravity, eqs.~(\ref{lcst}) and (\ref{fgl}), one of
the connections involved, namely $\omega$ or $\bar{\omega}$, is valued only on
the Lorentz subalgebra of the full AdS algebra. Here we consider some
implications of this fact.

Let us consider first the particular case of the Transgression form
$Q_{\bm{\omega}+\bm{e}\leftarrow\bm{\omega}}^{\left(  2n+1\right)  }$ with the
Levi-Civita tensor as $\mathfrak{so}\left(  2n,2\right)  $-invariant symmetric
polynomial,%
\begin{equation}
Q_{\bm{\omega}+\bm{e}\leftarrow\bm{\omega}}^{\left(  2n+1\right)  }=\left(
n+1\right)  \int_{0}^{1}dt\left\langle \bm{e}\left(  \bm{R}+t^{2}%
\bm{e}^{2}\right)  ^{n}\right\rangle .\label{qwew}%
\end{equation}

The {\small RHS} of eq.~(\ref{qwew}) has the remarkable feature of changing
only by a \emph{total derivative}~\footnote{As shown in Ref.~\cite{Zan00b},
this holds only for infinitesimal transformations; for finite transformations
(\ref{qwew}) changes by a closed form.} under the infinitesimal $\mathfrak{so}%
\left(  2n,2\right)  $ gauge transformations
\begin{align}
\delta e^{a}  &  =\lambda_{\;b}^{a}e^{b}-\text{D}_{\omega}\lambda
^{a},\label{ads1}\\
\delta\omega^{ab}  &  =-\text{D}_{\omega}\lambda^{ab}-\frac{1}{\ell^{2}%
}\left(  e^{a}\lambda^{b}-e^{b}\lambda^{a}\right)  .\label{ads2}%
\end{align}

This seems puzzling, because, as we saw in section~\ref{sym}, a Transgression
form in general should be fully invariant, with no additional terms appearing
under gauge transformations. This behavior has its origin in the fact that the
above transformations are gauge transformations for the connection
$\bm{A}=\bm{\omega}+\bm{e}$, but \emph{not} for the connection $\bar
{\bm{A}}=\bm{\omega}$. Furthermore, it looks impossible to simultaneously
define consistent gauge transformations for both connections
$\bm{A}=\bm{\omega}+\bm{e}$ and $\bar{\bm{A}}=\bm{\omega}$.

Bearing this in mind, it now seems amazing that (\ref{qwew}) changes only by a
closed term under these `pathological' gauge transformations! This puzzle is
related with some interesting properties of the choice of invariant symmetric tensor.

In order to shed some light onto this riddle, let us consider the
Transgression form $Q_{\bm{\omega}+\bm{e}\leftarrow\bm{\omega}}^{\left(
2n+1\right)  }$ with an arbitrary invariant symmetric tensor $\left\langle
\cdots\right\rangle $. Then, from the Chern--Weil theorem, we know that%
\begin{equation}
\left\langle \bm{F}^{n+1}\right\rangle -\left\langle \bm{R}^{n+1}\right\rangle
=\text{d}Q_{\bm{\omega}+\bm{e}\leftarrow\bm{\omega}}^{\left(  2n+1\right)
},\label{frq}%
\end{equation}
where $\bm{F}$ and $\bm{R}$ are the curvatures for $\bm{\omega}+\bm{e}$ and
$\bm{\omega}$, respectively.

Because the transformations (\ref{ads1})--(\ref{ads2}) are gauge
transformations for $\bm{A}=\bm{\omega}+\bm{e}$ but not for $\bar
{\bm{A}}=\bm{\omega}$, $\bm{F}$ will change under them as an $\mathfrak{so}%
\left(  2n,2\right)  $-tensor, but $\bm{R}$ will not. Therefore, $\left\langle
\bm{F}^{n+1}\right\rangle $ will stay invariant under (\ref{ads1}%
)--(\ref{ads2}) but not $\left\langle \bm{R}^{n+1}\right\rangle $. As a
consequence, the {\small LHS} in (\ref{frq}) will be modified under these
pathological gauge transformations, and then of course also the {\small RHS}.
In general, $Q_{\bm{\omega}+\bm{e}\leftarrow\bm{\omega}}^{\left(  2n+1\right)
}$ is simply not invariant at all under these transformations.

But when the invariant symmetric polynomial\ is such that $\left\langle
\bm{R}^{n+1}\right\rangle $ remains unchanged even under these non-gauge
transformations, then $Q_{\bm{\omega}+\bm{e}\leftarrow\bm{\omega}}^{\left(
2n+1\right)  }$ can at most vary by a closed form under them. This is
precisely the case for the choice of the Levi-Civita tensor. In this case,
because $\left\langle \bm{J}_{a_{1}a_{2}}\cdots\bm{J}_{a_{2n+1}a_{2n+2}%
}\right\rangle =0$, we have $\left\langle \bm{R}^{n+1}\right\rangle =0$ and
this value is not modified even under the `pathological' gauge transformations.

It is interesting to notice how remarkable the invariant symmetric polynomial
structure has been: it is also the reason why
$Q_{\bm{\omega }+\bm{e}\leftarrow\bm{\omega}}^{\left(  2n+1\right)  }$ and the
CS form differ only by a total derivative, and as we will see after some
discussion, it has further importance.

Setting aside the suggestive fact that $Q_{\bm{\omega}+\bm{e}\leftarrow
\bm{\omega}}^{\left(  2n+1\right)  }$ changes only by boundary contributions
under the transformations, the mathematical beauty problem remains: those are
not gauge transformations for our fields.

This problem has one clear solution: change the configuration of connections
into another one better behaved. This is precisely what happens with the
second expression used as Lagrangian, $Q_{\bm{\omega}+\bm{e}\leftarrow
\bar{\bm{\omega}}}^{\left(  2n+1\right)  }$. On this expression it is possible
to define in a consistent way simultaneous SO$\left(  2n,2\right)  $
\emph{gauge} transformations over $\bm{A}=\bm{\omega}+\bm{e}$ and
$\bar{\bm{A}}=\bm{\omega}$,%
\begin{align}
\delta\bm{A}  &  =-\frac{1}{2}\left[  \text{D}_{\omega}\lambda^{ab}+\frac
{1}{\ell^{2}}\left(  e^{a}\lambda^{b}-e^{b}\lambda^{a}\right)  \right]
\bm{J}_{ab}+\nonumber\\
&  +\left(  \lambda_{\;b}^{a}e^{b}-\text{D}_{\omega}\lambda^{a}\right)
\bm{P}_{a},\\
\delta\bar{\bm{A}}  &  =-\frac{1}{2}\text{D}_{\bar{\omega}}\lambda
^{ab}\bm{J}_{ab}-\text{D}_{\bar{\omega}}\lambda^{a}\bm{P}_{a},
\end{align}
such that $Q_{\bm{\omega}+\bm{e}\leftarrow\bar{\bm{\omega }}}^{\left(
2n+1\right)  }$ stays fully invariant under them, as it should be.

Two intriguing facts appear now. The first of them is that the difference
between $Q_{\bm{\omega}+\bm{e}\leftarrow\bar{\bm
{\omega}}}^{\left(  2n+1\right)  }$ and $Q_{\bm{\omega}+\bm{e}\leftarrow
\bm{\omega}}^{\left(  2n+1\right)  }$ is only a total derivative, and
therefore it seems surprising that the gauge transformations are ill-defined
over $Q_{\bm{\omega}+\bm{e}\leftarrow\bm{\omega}}^{\left(  2n+1\right)  }$,
whereas there is no problem with $Q_{\bm{\omega}+\bm{e}\leftarrow
\bar{\bm{\omega}}}^{\left(  2n+1\right)  }$. The second fact is that the
statement that $\bar{\bm{A}}=\bar{\bm{\omega}}$ has components only on the
Lorentz subalgebra is not a gauge-invariant one. As a matter of fact, an AdS
boost on $\bar{\bm{A}}$ will generate a $\bm{P}_{a}$-piece: a `pure-gauge'
vielbein (see section~\ref{lvsg}).

The first of these facts finds a natural explanation in the context of the
Triangle equation [cf. eq.~(\ref{treq})],%
\begin{equation}
Q_{\bm{A}_{2}\leftarrow\bm{A}_{0}}^{\left(  2n+1\right)  }=Q_{\bm{A}_{2}%
\leftarrow\bm{A}_{1}}^{\left(  2n+1\right)  }+Q_{\bm{A}_{1}\leftarrow
\bm{A}_{0}}^{\left(  2n+1\right)  }+\text{d}Q_{\bm{A}_{2}\leftarrow
\bm{A}_{1}\leftarrow\bm{A}_{0}}^{\left(  2n\right)  }.
\end{equation}

Here it suffices to notice that $Q_{\bm{A}_{2}\leftarrow\bm{A}_{0}}^{\left(
2n+1\right)  }$ does \emph{not} depend on $\bm{A}_{1}$, and it remains
invariant under any kind of transformation $\bm{A}_{1}\rightarrow
\bm{A}_{1}^{\prime}$, \emph{even when this is not a gauge transformation}.

For this reason, we can see that in the particular case $\bm{A}_{0}%
=\bar{\bm{\omega}}$, $\bm{A}_{1}=\bm{\omega}$, $\bm{A}_{2}=\bm{\omega}+\bm{e}$%
,
\[
Q_{\bm{\omega}+\bm{e}\leftarrow\bar{\bm{\omega}}}^{\left(  2n+1\right)
}=Q_{\bm{\omega}+\bm{e}\leftarrow\bm{\omega}}^{\left(  2n+1\right)
}+Q_{\bm{\omega}\leftarrow\bar{\bm{\omega}}}^{\left(  2n+1\right)  }%
+\text{d}Q_{\bm{\omega}+\bm{e}\leftarrow\bm{\omega}\leftarrow\bar
{\bm{\omega}}}^{\left(  2n\right)  },
\]
the Transgression form $Q_{\bm{\omega}+\bm{e}\leftarrow\bar{\bm{\omega}}%
}^{\left(  2n+1\right)  }$ will stay fully invariant under the infinitesimal
transformations
\begin{align}
\delta\bm{A}_{0}  &  =-\frac{1}{2}\text{D}_{\bar{\omega}}\lambda
^{ab}\bm{J}_{ab}-\text{D}_{\bar{\omega}}\lambda^{a}\bm{P}_{a},\\
\delta\bm{A}_{1}  &  =-\frac{1}{2}\left[  \text{D}_{\omega}\lambda^{ab}%
+\frac{1}{\ell^{2}}\left(  e^{a}\lambda^{b}-e^{b}\lambda^{a}\right)  \right]
\bm{J}_{ab},\\
\delta\bm{A}_{2}  &  =-\frac{1}{2}\left[  \text{D}_{\omega}\lambda^{ab}%
+\frac{1}{\ell^{2}}\left(  e^{a}\lambda^{b}-e^{b}\lambda^{a}\right)  \right]
\bm{J}_{ab}+\nonumber\\
&  +\left(  \lambda_{\;b}^{a}e^{b}-\text{D}_{\omega}\lambda^{a}\right)
\bm{P}_{a},
\end{align}
even though they do not correspond to a gauge transformation~\footnote{An
infinitesimal gauge transformation for $\bm{A}_{1}$ would rather read
$\delta\bm{A}_{1}=-\frac{1}{2}$D$_{\omega}\lambda^{ab}\bm{J}_{ab}-$D$_{\omega
}\lambda^{a}\bm{P}_{a}$.} for $\bm{A}_{1}$. The difference between the present
case and the former is that $Q_{\bm{\omega}+\bm{e}\leftarrow\bar{\bm{\omega}}%
}^{\left(  2n+1\right)  }$ does not depend on $\bm{A}_{1}$ and therefore
there's no contradiction in having non-gauge transformation laws for it.

On the other hand, when the Levi-Civita symbol is chosen, then
$Q_{\bm{\omega}\leftarrow\bar{\bm{\omega}}}^{\left(  2n+1\right)  }=0$ and we
can write
\begin{equation}
Q_{\bm{\omega}+\bm{e}\leftarrow\bar{\bm{\omega}}}^{\left(  2n+1\right)
}=Q_{\bm{\omega}+\bm{e}\leftarrow\bm{\omega}}^{\left(  2n+1\right)  }%
+\text{d}Q_{\bm{\omega }+\bm{e}\leftarrow\bm{\omega}\leftarrow\bar
{\bm {\omega}}}^{\left(  2n\right)  }.\label{Descomposition}%
\end{equation}

\subsection{\label{lvsg}Locality vs. Globality}

We would like to focus now our attention on a subtle and interesting fact
related to eq.~(\ref{Descomposition}): its {\small LHS} has more information
than its {\small RHS}. As matter of fact, under an $\mathfrak{so}\left(
2n,2\right)  $ gauge transformation, $\bm{A}_{0}=\bar{\bm {\omega}}$ changes
as
\begin{equation}
\bm{A}_{0}=\bar{\bm{\omega}}\rightarrow\bm{A}_{0}^{\prime}=\bar{\bm{\omega}}%
^{\prime}+\bar{\bm{e}}_{\text{g}},\label{pgv}%
\end{equation}
where $\bar{\bm{e}}_{\text{g}}$ is a `pure-gauge' vielbein. In this situation,
even though $Q_{\bm{\omega}\leftarrow\bar{\bm
{\omega}}}^{\left(  2n+1\right)  }$ is zero, \emph{its variation under
(\ref{pgv}) does not vanish}, i.e.,
\begin{equation}
Q_{\bm{\omega}\leftarrow\bar{\bm{\omega}}}^{\left(  2n+1\right)
}=0\rightarrow Q_{\bm{\omega}^{\prime}\leftarrow\bar{\bm{\omega}}^{\prime
}+\bar{\bm{e}}_{\text{g}}}^{\left(  2n+1\right)  }\neq0.
\end{equation}
Therefore, despite the fact that $Q_{\bm{\omega}+\bm{e}\leftarrow
\bar{\bm{\omega}}}^{\left(  2n+1\right)  }$ is fully gauge invariant, if we
perform a gauge transformation over just $\tilde{L}%
_{\bm{\omega}+\bm{e}\leftarrow\bar{\bm{\omega}}}^{\left(  2n+1\right)  }\equiv
k\left(  Q_{\bm{\omega}+\bm{e}\leftarrow\bm{\omega}}^{\left(  2n+1\right)
}+\text{d}Q_{\bm{\omega}+\bm{e}\leftarrow\bm{\omega}\leftarrow\bar
{\bm{\omega}}}^{\left(  2n\right)  }\right)  $, the result is%
\begin{equation}
\tilde{L}_{\bm{\omega}+\bm{e}\leftarrow\bar{\bm {\omega}}}^{\left(
2n+1\right)  }\rightarrow\tilde{L}_{\bm{\omega }^{\prime}+\bm{e}^{\prime
}\leftarrow\bar{\bm{\omega}}^{\prime}}^{\left(  2n+1\right)  }=\tilde
{L}_{\bm{\omega}+\bm{e}\leftarrow\bar{\bm{\omega}}}^{\left(  2n+1\right)
}-kQ_{\bm{\omega}^{\prime}\leftarrow\bar{\bm{\omega}}^{\prime}+\bar
{\bm{e}}_{\text{g}}}^{\left(  2n+1\right)  }.
\end{equation}

It is once again only due to the very special properties of the Levi-Civita
tensor that $Q_{\bm{\omega}^{\prime}\leftarrow\bar{\bm{\omega
}}^{\prime}+\bar{\bm{e}}_{\text{g}}}^{\left(  2n+1\right)  }$ can be shown to
be at most a closed form. In this way, we can observe that
$L_{\bm{\omega}+\bm{e}\leftarrow\bar{\bm{\omega}}}^{\left(  2n+1\right)
}=kQ_{\bm{\omega}+\bm{e}\leftarrow\bar{\bm{\omega}}}^{\left(  2n+1\right)  }$
is a fully gauge-invariant, globally-defined expression for the Lagrangian,
but that on the other hand $\tilde{L}_{\bm{\omega}+\bm{e}\leftarrow
\bar{\bm{\omega}}}^{\left(  2n+1\right)  }=k\left(
Q_{\bm{\omega}+\bm{e}\leftarrow\bm{\omega}}^{\left(  2n+1\right)  }%
+\text{d}Q_{\bm{\omega}+\bm{e}\leftarrow\bm{\omega}\leftarrow\bar
{\bm{\omega}}}^{\left(  2n\right)  }\right)  $ is an easier to evaluate
expression, which describes the dynamics of the theory but holds only locally.

It may seem we have been abusing the \textquotedblleft$=$\textquotedblright%
\ symbol. Saving it for equalities which are preserved under gauge
transformations and using instead \textquotedblleft$\approx$\textquotedblright%
\ for the ones which are not, we may write $Q_{\bm{\omega}\leftarrow
\bar{\bm{\omega}}}^{\left(  2n+1\right)  }\approx0$ and
\begin{align}
L_{\bm{\omega}+\bm{e}\leftarrow\bar{\bm{\omega}}}^{\left(  2n+1\right)  }  &
=kQ_{\bm{\omega}+\bm{e}\leftarrow\bar{\bm{\omega}}}^{\left(  2n+1\right)
},\nonumber\\
&  \approx k\left(  Q_{\bm{\omega}+\bm{e}\leftarrow\bm{\omega}}^{\left(
2n+1\right)  }+\text{d}Q_{\bm{\omega }+\bm{e}\leftarrow\bm{\omega}\leftarrow
\bar{\bm {\omega}}}^{\left(  2n\right)  }\right)  .
\end{align}

\subsection{Theory Doubling}

Under the light of all the above discussion, and seeking just mathematical
beauty and symmetry, it may seem interesting to consider the lagrangian%
\begin{equation}
L_{\bm{\omega}+\bm{e}\leftarrow\bar{\bm{\omega}}+\bar{\bm{e}}}^{\left(
2n+1\right)  }=kQ_{\bm{\omega }+\bm{e}\leftarrow\bar{\bm{\omega}}+\bar
{\bm{e}}}^{\left(  2n+1\right)  }.
\end{equation}

Using the Triangle equation (\ref{treq}), we get
\begin{align}
L_{\bm{\omega}+\bm{e}\leftarrow\bar{\bm{\omega}}+\bar{\bm{e}}}^{\left(
2n+1\right)  }  &  =kQ_{\bm{\omega }+\bm{e}\leftarrow\bm{\omega}}^{\left(
2n+1\right)  }-kQ_{\bar{\bm{\omega}}+\bar{\bm{e}}\leftarrow\bar{\bm {\omega}}%
}^{\left(  2n+1\right)  }+kQ_{\bm{\omega}\leftarrow\bar{\bm{\omega}}}^{\left(
2n+1\right)  }+\nonumber\\
&  +k\text{d}\left(  Q_{\bm{\omega}+\bm{e}\leftarrow\bm{\omega}\leftarrow
\bar{\bm{\omega}}}^{\left(  2n\right)  }+Q_{\bm{\omega}+\bm{e}\leftarrow
\bar{\bm{\omega}}\leftarrow\bar{\bm{\omega}}+\bar{\bm{e}}}^{\left(  2n\right)
}\right)  .
\end{align}

The ECHF (\ref{cehf}) with $p=2$ may now be used to yield%

\begin{align}
\text{d}Q_{\bm{A}_{3}\leftarrow\bm{A}_{2}\leftarrow\bm{A}_{1}\leftarrow
\bm{A}_{0}}^{\left(  2n-1\right)  }  &  =Q_{\bm{A}_{3}\leftarrow
\bm{A}_{2}\leftarrow\bm{A}_{1}}^{\left(  2n\right)  }-Q_{\bm{A}_{3}%
\leftarrow\bm{A}_{2}\leftarrow\bm{A}_{0}}^{\left(  2n\right)  }+\nonumber\\
&  +Q_{\bm{A}_{3}\leftarrow\bm{A}_{1}\leftarrow\bm{A}_{0}}^{\left(  2n\right)
}-Q_{\bm{A}_{2}\leftarrow\bm{A}_{1}\leftarrow\bm{A}_{0}}^{\left(  2n\right)
},
\end{align}
where this time $Q_{\bm{A}_{3}\leftarrow\bm{A}_{2}\leftarrow\bm{A}_{1}%
\leftarrow\bm{A}_{0}}^{\left(  2n-1\right)  }$ is a $\left(  2n-1\right)
$-form on the space-time manifold $M$ which is integrated over the 3-simplex.
Plugging in our connections, it is possible to write down the Lagrangian as%
\begin{align}
L_{\bm{\omega}+\bm{e}\leftarrow\bar{\bm{\omega}}+\bar{\bm{e}}}^{\left(
2n+1\right)  }  &  =kQ_{\bm{\omega }+\bm{e}\leftarrow\bm{\omega}}^{\left(
2n+1\right)  }-kQ_{\bar{\bm{\omega}}+\bar{\bm{e}}\leftarrow\bar{\bm {\omega}}%
}^{\left(  2n+1\right)  }+kQ_{\bm{\omega}\leftarrow\bar{\bm{\omega}}}^{\left(
2n+1\right)  }+\nonumber\\
&  +\frac{1}{2}k\text{d}\left(  Q_{\bm{\omega}+\bm{e}\leftarrow
\bm{\omega}\leftarrow\bar{\bm{\omega}}}^{\left(  2n\right)  }%
+Q_{\bm{\omega}\leftarrow\bar{\bm{\omega}}\leftarrow\bar{\bm{\omega}}%
+\bar{\bm{e}}}^{\left(  2n\right)  }+\right. \nonumber\\
&  \left.  +Q_{\bm{\omega}+\bm{e}\leftarrow\bar{\bm{\omega}}\leftarrow
\bar{\bm{\omega}}+\bar{\bm{e}}}^{\left(  2n\right)  }%
+Q_{\bm{\omega}+\bm{e}\leftarrow\bm{\omega}\leftarrow\bar{\bm{\omega}}%
+\bar{\bm{e}}}^{\left(  2n\right)  }\right)  .
\end{align}

This way of writing the lagrangian allows us to see the completely symmetrical
r\^{o}le that the $\bm{\omega}+\bm{e}\ $and $\bar{\bm
{\omega}}+\bar{\bm{e}}$ connections play within it. When the Levi-Civita
symbol is chosen, $L_{\bm{\omega}+\bm{e}\leftarrow\bar{\bm{\omega}}%
+\bar{\bm{e}}}^{\left(  2n+1\right)  }$ tells us about two identical,
independent LL Gravity theories in the bulk that interact only at the boundary.

One important aspect of the $L_{\bm{\omega}+\bm{e}\leftarrow\bar
{\bm{\omega}}+\bar{\bm{e}}}^{\left(  2n+1\right)  }$ lagrangian concerns its
transformation properties under parity and time inversion. Under a PT
transformation, $L_{\bm{\omega
}+\bm{e}\leftarrow\bar{\bm{\omega}}+\bar{\bm{e}}}^{\left(  2n+1\right)  }$
flips sign. This means that, if we rather na\"{\i}vely interpret the
interchange $\bm{A}\leftrightarrows\bar{\bm{A}}$ as charge conjugation C (see
section~\ref{carg}), then $L_{\bm{\omega}+\bm{e}\leftarrow\bar{\bm{\omega}}%
+\bar{\bm{e}}}^{\left(  2n+1\right)  }$ turns out to be invariant under the
combined CPT operation. Even though this bears some resemblance of a
particle-antiparticle relation, where the only interaction would occur on
the space-time boundary, more work is clearly needed in order to fully solve
this issue.

It is interesting to observe that this theory doubling is quite general, and
not only privative of gravity. From the Triangle equation, and fixing the
middle connection to zero, it is possible to observe that%
\begin{align}
L_{\bm{A}\leftarrow\bar{\bm{A}}}^{\left(  2n+1\right)  }  &
=kQ_{\bm{A}\leftarrow\bar{\bm{A}}}^{\left(  2n+1\right)  },\nonumber\\
&  =kQ_{\bm{A}\leftarrow0}^{\left(  2n+1\right)  }-kQ_{\bar{\bm{A}}%
\leftarrow0}^{\left(  2n+1\right)  }+k\text{d}Q_{\bm{A}\leftarrow
0\leftarrow\bar{\bm{A}}}^{\left(  2n\right)  },\nonumber\\
&  =L_{\text{CS}}^{\left(  2n+1\right)  }\left(  \bm{A}\right)  -L_{\text{CS}%
}^{\left(  2n+1\right)  }\left(  \bar{\bm{A}}\right)  +k\text{d}%
Q_{\bm{A}\leftarrow0\leftarrow\bar{\bm{A}}}^{\left(  2n\right)  }.
\end{align}

In this way we see that this behaviour\ can arise in any kind of CS theory,
for example in CS SUGRA.

Despite of the mathematical appeal this kind of symmetrical double structure
possesses, its physical interpretation seems a bit unclear. It is for this
reason that lagrangians such as $L_{\bm{\omega}+\bm{e}\leftarrow
\bar{\bm{\omega}}}^{\left(  2n+1\right)  }=kQ_{\bm{\omega}+\bm{e}\leftarrow
\bar{\bm{\omega}}}^{\left(  2n+1\right)  }$ are so interesting. In it, the
connection $\bar{\bm{\omega}}$ enters only through the boundary term as the
connection associated to the intrinsic curvature of the boundary, and this
looks a lot more satisfactory from a physical point of view.

We have been amazed by the fact that the awkward-looking lagrangian
$L_{\bm{\omega}+\bm{e}\leftarrow\bar{\bm{\omega}}+\bar{\bm{e}}}^{\left(
2n+1\right)  }$ reduces to the physically sensible
$L_{\bm{\omega}+\bm{e}\leftarrow\bar{\bm
{\omega}}}^{\left(  2n+1\right)  }$ when $\bar{\bm{e}}$ corresponds to pure
gauge. As shown in Appendix~\ref{apx}, a pure gauge vielbein turns out to be
as sensible an idea as flat space-time itself.

\begin{acknowledgments}
The authors wish to thank J.~A.~de~Azc\'{a}rraga for his warm hospitality at
the Universitat de Val\`{e}ncia, where this work was brought to completion.
F.~I. and E.~R. wish to thank D.~L\"{u}st for his kind hospitality at the
Humboldt-Universit\"{a}t zu Berlin and at the Arnold Sommerfeld Center for
Theoretical Physics in Munich, and J.~Zanelli for having directed their
attention towards this problem through many enlightening discussions. The
research of P.~S. was partially supported by FONDECYT Grants 1040624, 1051086
and by Universidad de Concepci\'{o}n through Semilla Grants 205.011.036-1S,
205.011.037-1S. The research of F.~I. and E.~R. was supported by Ministerio de
Educaci\'{o}n through MECESUP Grant UCO 0209 and by the German Academic
Exchange Service (DAAD).

\textit{Note added}: After posting this paper to the arXiv we have been
informed of some results which partially overlap with those obtained in the
present work. We would like to thank A.~Borowiec for pointing out to us
Refs.~\cite{Bor03,Bor05} as well as J.~Zanelli for acquainting us with
P.~Mora's work~\cite{MorTh,Mor00,Mor01}, which we warmly acknowledge.
\end{acknowledgments}

\appendix

\section{\label{prcw}On the Proof of the Chern--Weil Theorem}

In this Appendix we provide a physicist's sketch of a proof for the
Chern--Weil theorem. We refer the reader to the literature for the required
mathematical rigor lacking in the following lines.

The proof uses the following identity [which is straightforward to establish
from definitions (\ref{thk})--(\ref{ft})]:
\begin{equation}
\frac{d}{dt}\bm{F}_{t}=\text{D}_{t}\bm{\theta},\label{dfdt}%
\end{equation}
where D$_{t}$ stands for the covariant derivative in the\ connection
$\bm{A}_{t}$.

We start by writing the {\small LHS} of (\ref{cwt}) as
\[
\left\langle \bm{F}^{n+1}\right\rangle -\left\langle \bar{\bm{F}}%
^{n+1}\right\rangle =\int_{0}^{1}dt\frac{d}{dt}\left\langle \bm{F}_{t}%
^{n+1}\right\rangle ,
\]
and then proceed with Leibniz's rule:
\[
\left\langle \bm{F}^{n+1}\right\rangle -\left\langle \bar{\bm{F}}%
^{n+1}\right\rangle =\left(  n+1\right)  \int_{0}^{1}dt\left\langle \frac
{d}{dt}\bm{F}_{t}\bm{F}_{t}^{n}\right\rangle .
\]
Applying (\ref{dfdt}), we get
\[
\left\langle \bm{F}^{n+1}\right\rangle -\left\langle \bar{\bm{F}}%
^{n+1}\right\rangle =\left(  n+1\right)  \int_{0}^{1}dt\left\langle
\text{D}_{t}\bm{\theta F}_{t}^{n}\right\rangle .
\]
Bianchi's identity D$_{t}\bm{F}_{t}=0$ now leads us to
\[
\left\langle \bm{F}^{n+1}\right\rangle -\left\langle \bar{\bm{F}}%
^{n+1}\right\rangle =\left(  n+1\right)  \int_{0}^{1}dt\left\langle
\text{D}_{t}\left(  \bm{\theta F}_{t}^{n}\right)  \right\rangle ,
\]
and the invariance property of $\left\langle \cdots\right\rangle $ further
implies that
\[
\left\langle \bm{F}^{n+1}\right\rangle -\left\langle \bar{\bm{F}}%
^{n+1}\right\rangle =\left(  n+1\right)  \text{d}\int_{0}^{1}dt\left\langle
\bm{\theta F}_{t}^{n}\right\rangle ,
\]
which completes the proof.

\section{\label{feqbc}Derivation of the TGFT Field Equations and Boundary
Conditions}

In this Appendix we present the derivation of the TGFT field equations and
boundary conditions from the TGFT action, eq. (\ref{st}). The remarkable
simplicity and elegance of this derivation provide a striking proof of the
power of the TGFT formalism, as made evident in the following lines.

We begin with the TGFT Lagrangian,
\begin{equation}
L_{\bm{A}\leftarrow\bar{\bm{A}}}^{\left(  2n+1\right)  }=\left(  n+1\right)
k\int_{0}^{1}dt\left\langle \bm{\theta F}_{t}^{n}\right\rangle ,\label{lt4}%
\end{equation}
where
\begin{align}
\bm{\theta}  &  =\bm{A}-\bar{\bm{A}},\\
\bm{A}_{t}  &  =\bar{\bm{A}}+t\bm{\theta}.
\end{align}
Under the independent infinitesimal variations $\bm{A}\rightarrow
\bm{A}+\delta\bm{A}$, $\bar{\bm{A}}\rightarrow\bar{\bm{A}}+\delta\bar{\bm{A}}%
$, $\bm{\theta}$ and $\bm{A}_{t}$ change by $\delta\bm{\theta}=\delta
\bm{A}-\delta\bar{\bm{A}}$, $\delta\bm{A}_{t}=\delta\bar{\bm{A}}%
+t\delta\bm{\theta} $ and the lagrangian varies by
\begin{align}
\delta L_{\bm{A}\leftarrow\bar{\bm{A}}}^{\left(  2n+1\right)  }  &  =n\left(
n+1\right)  k\int_{0}^{1}dt\left\langle \bm{\theta}\text{D}_{t}\delta
\bm{A}_{t}\bm{F}_{t}^{n-1}\right\rangle +\nonumber\\
&  +\left(  n+1\right)  k\int_{0}^{1}dt\left\langle \delta\bm{\theta F}_{t}%
^{n}\right\rangle ,\label{v4}%
\end{align}
where we have used the well-known identity $\delta\bm{F}_{t}=$ D$_{t}%
\delta\bm{A}_{t}$.

Leibniz's rule for D$_{t}$ and the invariance property of the symmetric
polynomial $\left\langle \cdots\right\rangle $ now allow us to write
\begin{equation}
\left\langle \bm{\theta}\text{D}_{t}\delta\bm{A}_{t}\bm{F}_{t}^{n-1}%
\right\rangle =\left\langle \text{D}_{t}\bm{\theta}\delta\bm{A}_{t}%
\bm{F}_{t}^{n-1}\right\rangle -\text{d}\left\langle \bm{\theta}\delta
\bm{A}_{t}\bm{F}_{t}^{n-1}\right\rangle .
\end{equation}

From the identities
\begin{align}
\frac{d}{dt}\bm{F}_{t}  &  =\text{D}_{t}\bm{\theta},\\
\frac{d}{dt}\delta\bm{A}_{t}  &  =\delta\bm{\theta},
\end{align}
and Leibniz's rule for $d/dt$, it follows that
\begin{align}
n\left\langle \bm{\theta}\text{D}_{t}\delta\bm{A}_{t}\bm{F}_{t}^{n-1}%
\right\rangle  &  =\frac{d}{dt}\left\langle \delta\bm{A}_{t}\bm{F}_{t}%
^{n}\right\rangle -\left\langle \delta\bm{\theta F}_{t}^{n}\right\rangle
+\nonumber\\
&  -n\text{d}\left\langle \bm{\theta}\delta\bm{A}_{t}\bm{F}_{t}^{n-1}%
\right\rangle .\label{n4}%
\end{align}

Plugging (\ref{n4}) in (\ref{v4}) we get%

\begin{align}
\delta L_{\bm{A}\leftarrow\bar{\bm{A}}}^{\left(  2n+1\right)  }  &  =\left(
n+1\right)  k\int_{0}^{1}dt\frac{d}{dt}\left\langle \delta\bm{A}_{t}%
\bm{F}_{t}^{n}\right\rangle +\nonumber\\
&  +n\left(  n+1\right)  k\text{d}\int_{0}^{1}dt\left\langle \delta
\bm{A}_{t}\bm{\theta F}_{t}^{n-1}\right\rangle ,
\end{align}
which leads us directly to our final result:%

\begin{align}
\delta L_{\bm{A}\leftarrow\bar{\bm{A}}}^{\left(  2n+1\right)  }  &  =\left(
n+1\right)  k\left(  \left\langle \delta\bm{AF}^{n}\right\rangle -\left\langle
\delta\bar{\bm{A}\bar{F}}^{n}\right\rangle \right)  +\nonumber\\
&  +n\left(  n+1\right)  k\text{d}\int_{0}^{1}dt\left\langle \delta
\bm{A}_{t}\bm{\theta F}_{t}^{n-1}\right\rangle .
\end{align}

\section{\label{ccs}Conserved Currents for the TGFT Action}

A derivation of the Noether currents for the TGFT action is presented. As with
the field equations and boundary conditions, the power of the TGFT formalism
becomes evident in the simplicity of the following derivation.

With the notation of section~\ref{carg}, we have [cf. eq. (\ref{th})]%
\begin{equation}
\Theta\left(  \bm{A},\bar{\bm{A}},\delta\bm{A},\delta\bar{\bm{A}}\right)
=n\left(  n+1\right)  k\int_{0}^{1}dt\left\langle \delta\bm{A}_{t}%
\bm{\theta F}_{t}^{n-1}\right\rangle ,\label{th5}%
\end{equation}
and the conserved Noether currents%
\begin{align}
\left.  \star J_{\text{gauge}}\right.   &  =-\Theta\left(  \bm{A},\bar
{\bm{A}},-\text{D}_{\bm{A}}\bm{\lambda},-\text{D}_{\bar{\bm{A}}}%
\bm{\lambda}\right)  ,\label{ga55}\\
\left.  \star J_{\text{diff}}\right.   &  =-\Theta\left(  \bm{A},\bar
{\bm{A}},-\pounds _{\xi}\bm{A},-\pounds _{\xi}\bar{\bm{A}}\right)
-\text{I}_{\xi}L_{\bm{A}\leftarrow\bar{\bm{A}}}^{\left(  2n+1\right)
}.\label{df5}%
\end{align}

\subsection{Conserved current for gauge transformations}

Let us start with the gauge conserved current. Replacing (\ref{th5}) in
(\ref{ga55}), we have
\begin{equation}
\left.  \star J_{\text{gauge}}\right.  =-n\left(  n+1\right)  k\int_{0}%
^{1}dt\left\langle \bm{\theta}\text{D}_{t}\bm{\bm{\lambda}F}_{t}%
^{n-1}\right\rangle .
\end{equation}

Using Bianchi's identity D$_{t}\bm{F}_{t}=0$, Leibniz's rule for D$_{t}$ and
the invariance property of the symmetric polynomial $\left\langle
\cdots\right\rangle $, we get
\begin{align}
\left.  \star J_{\text{gauge}}\right.   &  =n\left(  n+1\right)  k\text{d}%
\int_{0}^{1}dt\left\langle \bm{\theta\lambda F}_{t}^{n-1}\right\rangle
+\nonumber\\
&  -n\left(  n+1\right)  k\int_{0}^{1}dt\left\langle \bm{\lambda}\text{D}%
_{t}\bm{\theta F}_{t}^{n-1}\right\rangle .
\end{align}

Now we replace the identity $\left(  d/dt\right)  \bm{F}_{t}=$ D$_{t}%
\bm{\theta}$ and integrate by parts in $t$ to obtain
\begin{align}
\left.  \star J_{\text{gauge}}\right.   &  =n\left(  n+1\right)  k\text{d}%
\int_{0}^{1}dt\left\langle \bm{\theta\lambda F}_{t}^{n-1}\right\rangle
+\nonumber\\
&  -\left(  n+1\right)  k\int_{0}^{1}dt\frac{d}{dt}\left\langle
\bm{\lambda F}_{t}^{n}\right\rangle ,
\end{align}
which leads us directly to
\begin{align}
\left.  \star J_{\text{gauge}}\right.   &  =n\left(  n+1\right)  k\text{d}%
\int_{0}^{1}dt\left\langle \bm{\theta\lambda F}_{t}^{n-1}\right\rangle
+\nonumber\\
&  -\left(  n+1\right)  k\left(  \left\langle \bm{\lambda F}^{n}\right\rangle
-\left\langle \bm{\lambda}\bar{\bm{F}}^{n}\right\rangle \right)  .\label{lt5}%
\end{align}

Since the last term in (\ref{lt5}) is proportional to the equations of motion,
we finally get the on-shell conserved Noether current
\begin{equation}
\left.  \star J_{\text{gauge}}\right.  =n\left(  n+1\right)  k\text{d}\int
_{0}^{1}dt\left\langle \bm{\theta\lambda F}_{t}^{n-1}\right\rangle .
\end{equation}

\subsection{Conserved current for diffeomorphisms}

Let us now consider the diffeomorphism conserved current. Replacing
(\ref{th5}) in (\ref{df5}) we get the following expression for the Noether
current:
\begin{equation}
\left.  \star J_{\text{diff}}\right.  =-n\left(  n+1\right)  k\int_{0}%
^{1}dt\left\langle \bm{\theta}\pounds _{\xi}\bm{A}_{t}\bm{F}_{t}%
^{n-1}\right\rangle -\text{I}_{\xi}L_{\bm{A}\leftarrow\bar{\bm{A}}}^{\left(
2n+1\right)  }.
\end{equation}

The identity%
\begin{equation}
\pounds _{\xi}\bm{A}_{t}=\text{I}_{\xi}\bm{F}_{t}+\text{D}_{t}\text{I}_{\xi
}\bm{A}_{t}%
\end{equation}
can be used to yield
\begin{align}
\left.  \star J_{\text{diff}}\right.   &  =-n\left(  n+1\right)  k\int_{0}%
^{1}dt\left\langle \bm{\theta}\text{I}_{\xi}\bm{F}_{t}\bm{F}_{t}%
^{n-1}\right\rangle +\nonumber\\
&  -n\left(  n+1\right)  k\int_{0}^{1}dt\left\langle \bm{\theta}\text{D}%
_{t}\text{I}_{\xi}\bm{A}_{t}\bm{F}_{t}^{n-1}\right\rangle -\text{I}_{\xi
}L_{\bm{A}\leftarrow\bar{\bm{A}}}^{\left(  2n+1\right)  }.
\end{align}

Using Leibniz's rule for both the contraction operator I$_{\xi}$ and the
covariant derivative D$_{t}$, plus Bianchi's identity D$_{t}\bm{F}_{t}=0$ and
the invariant nature of the symmetric polynomial $\left\langle \cdots
\right\rangle $, we find the expression%
\begin{align}
\left.  \star J_{\text{diff}}\right.   &  =-\left(  n+1\right)  k\int_{0}%
^{1}dt\left\langle \bm{\theta}\text{I}_{\xi}\bm{F}_{t}^{n}\right\rangle
+\nonumber\\
&  +n\left(  n+1\right)  k\text{d}\int_{0}^{1}dt\left\langle
\bm{\theta}\text{I}_{\xi}\bm{A}_{t}\bm{F}_{t}^{n-1}\right\rangle +\nonumber\\
&  -n\left(  n+1\right)  k\int_{0}^{1}dt\left\langle \text{I}_{\xi}%
\bm{A}_{t}\text{D}_{t}\bm{\theta F}_{t}^{n-1}\right\rangle -\text{I}_{\xi
}L_{\bm{A}\leftarrow\bar{\bm{A}}}^{\left(  2n+1\right)  }.
\end{align}

Now we make use of the identities
\begin{align}
\frac{d}{dt}\bm{F}_{t}  &  =\text{D}_{t}\bm{\theta},\\
\frac{d}{dt}\text{I}_{\xi}\bm{A}_{t}  &  =\text{I}_{\xi}\bm{\theta},\nonumber
\end{align}
and integrate by parts in $t$ to get
\begin{align}
\left.  \star J_{\text{diff}}\right.   &  =-\left(  n+1\right)  k\int_{0}%
^{1}dt\left\langle \bm{\theta}\text{I}_{\xi}\bm{F}_{t}^{n}\right\rangle
+\nonumber\\
&  +n\left(  n+1\right)  k\text{d}\int_{0}^{1}dt\left\langle
\bm{\theta}\text{I}_{\xi}\bm{A}_{t}\bm{F}_{t}^{n-1}\right\rangle +\nonumber\\
&  -\left(  n+1\right)  k\int_{0}^{1}dt\frac{d}{dt}\left\langle \text{I}_{\xi
}\bm{A}_{t}\bm{F}_{t}^{n}\right\rangle +\nonumber\\
&  +\left(  n+1\right)  k\int_{0}^{1}dt\left\langle \text{I}_{\xi
}\bm{\theta F}_{t}^{n}\right\rangle -\text{I}_{\xi}L_{\bm{A}\leftarrow
\bar{\bm {A}}}^{\left(  2n+1\right)  }.
\end{align}

Replacing the explicit form of the TGFT lagrangian in the last term we find
\begin{align}
\text{I}_{\xi}L_{\bm{A}\leftarrow\bar{\bm{A}}}^{\left(  2n+1\right)  }  &
=\left(  n+1\right)  k\int_{0}^{1}dt\left\langle \text{I}_{\xi}%
\bm{\theta F}_{t}^{n}\right\rangle +\nonumber\\
&  -\left(  n+1\right)  k\int_{0}^{1}dt\left\langle \bm{\theta}\text{I}_{\xi
}\bm{F}_{t}^{n}\right\rangle ,
\end{align}
so that%
\begin{align}
\left.  \star J_{\text{diff}}\right.   &  =n\left(  n+1\right)  k\text{d}%
\int_{0}^{1}dt\left\langle \bm{\theta}\text{I}_{\xi}\bm{A}_{t}\bm{F}_{t}%
^{n-1}\right\rangle +\nonumber\\
&  -\left(  n+1\right)  k\left(  \left\langle \text{I}_{\xi}\bm{AF}^{n}%
\right\rangle -\left\langle \text{I}_{\xi}\bar{\bm{A}\bar{F}}^{n}\right\rangle
\right)  .\label{la5}%
\end{align}

As with the gauge current, the last term in (\ref{la5}) is proportional to the
equations of motion. We finally obtain the on-shell conserved Noether current
\begin{equation}
\left.  \star J_{\text{diff}}\right.  =n\left(  n+1\right)  k\text{d}\int
_{0}^{1}dt\left\langle \bm{\theta}\text{I}_{\xi}\bm{A}_{t}\bm{F}_{t}%
^{n-1}\right\rangle .
\end{equation}

\section{\label{apx}Pure-Gauge Vielbeine}

The construction analyzed in this paper has a number of perhaps trivial but
quite interesting and beautiful features, which deserve to be explicitly mentioned.

One first simple fact with far-reaching consequences is that we are
considering \emph{both}, the spin connection \emph{and} the vielbein as
components of an $\mathfrak{so}\left(  2n,2\right)  $-valued, one-form gauge
connection
\begin{equation}
\bm{A}=\bm{\omega}+\bm{e}.
\end{equation}
This is a beautiful mathematical construction from a purely geometrical point
of view, because it unifies two different geometrical concepts, parallelism
and metricity, in a unique fiber bundle structure (see Refs.
\cite{Zan02,Zan05}). On the other hand, it is a key feature in the
construction of a background-free theory, because metricity enters in the
action on the same footing as all the other fields in the game.

We will now focus our attention on transformations generated by $g=\exp\left(
\zeta^{a}\bm{P}_{a}\right)  $, since they mix the vielbein and the spin
connection. Under a finite boost, $\bm{A}=\bm{\omega }+\bm{e}$ changes to
\begin{equation}
\bm{\omega}+\bm{e}\rightarrow\bm{\omega}^{\prime}+\bm{e}^{\prime}=g\left(
\text{d}+\bm{\omega}+\bm{e}\right)  g^{-1}.
\end{equation}
An explicit computation~\cite{Iza04,Sal03,Ste80}\ yields
\begin{align}
e^{\prime a}  &  =\Omega_{\;b}^{a}\left(  \cosh z\right)  e^{b}-\Omega
_{\;b}^{a}\left(  \frac{\sinh z}{z}\right)  \text{D}_{\omega}\zeta^{b},\\
\omega^{\prime ab}  &  =\omega^{ab}+\frac{1}{\ell^{2}}\left[  \frac{\sinh
z}{z}\left(  \zeta^{a}e^{b}-\zeta^{b}e^{a}\right)  +\right. \nonumber\\
&  \left.  \left(  \frac{1-\cosh z}{z^{2}}\right)  \left(  \zeta^{a}%
\text{D}_{\omega}\zeta^{b}-\zeta^{b}\text{D}_{\omega}\zeta^{a}\right)
\right]  ,
\end{align}
where
\begin{align}
\Omega_{\;b}^{a}\left(  u\right)   &  \equiv u\delta_{b}^{a}+\left(
1-u\right)  \frac{\zeta^{a}\zeta_{b}}{\zeta^{2}},\\
z  &  \equiv\frac{1}{\ell}\left(  \zeta^{a}\zeta_{a}\right)  ^{1/2}.
\end{align}
Our goal now is to find a geometry where parallelism is arbitrary but
metricity is pure gauge; i.e., a field configuration which differs from the
case $\bm{\omega}$ arbitrary and $\bm{e}=0$ only by a gauge transformation.

As a first step towards this goal, we consider a connection obtained from
gauging the simple case $\bm{A}=\bm{\omega}$,
\begin{align}
e^{\prime a} &  =-\Omega_{\;b}^{a}\left(  \frac{\sinh z}{z}\right)
\text{D}_{\omega}\zeta^{b},\label{ep}\\
\omega^{\prime ab} &  =\omega^{ab}+\frac{1}{\ell^{2}}\left(  \frac{1-\cosh
z}{z^{2}}\right)  \left(  \zeta^{a}\text{D}_{\omega}\zeta^{b}-\zeta
^{b}\text{D}_{\omega}\zeta^{a}\right)  .\label{wp}%
\end{align}
Eq.~(\ref{ep}) gives us $\bm{e}^{\prime}=\bm{e}^{\prime}\left(
\bm{\omega}\right)  $ and not $\bm{e}^{\prime}=\bm{e}^{\prime}\left(
\bm{\omega}^{\prime}\right)  $, which is what we need. The solution to this
problem involves performing the inverse gauge transformation to obtain
$\bm{\omega}=\bm{\omega}\left(  \bm{e}^{\prime},\bm{\omega}^{\prime}\right)
$. After some algebra, we find the connection%
\begin{equation}
\bm{A}=\bm{\omega}+\bm{e}_{\text{g}},\label{apg}%
\end{equation}
where
\begin{equation}
e_{\text{g}}^{a}=-\Omega_{\;b}^{a}\left(  \frac{\tanh z}{z}\right)
\text{D}_{\omega}\zeta^{b}%
\end{equation}
corresponds to a pure-gauge vielbein. The transformation given by
$g=\exp\left(  -\zeta^{a}\bm{P}_{a}\right)  $ takes (\ref{apg}) into one where
the vielbein vanishes and metricity as such ceases to exist.

This lack of metricity may seem like an extremely pathological feature to the
reader, but we would like to demystify it a little bit if we can.

We don't really need to go very far to find an example, as a simple one is
provided by AdS space itself, where we have
\begin{equation}
\bm{F}=\frac{1}{2}\left(  R^{ab}+\frac{1}{\ell^{2}}e^{a}e^{b}\right)
\bm{J}_{ab}+T^{a}\bm{P}_{a}=0.
\end{equation}
Even simpler is the case of Minkowski space, which can be obtained by
performing the \.{I}n\"{o}n\"{u}--Wigner contraction $\ell\rightarrow\infty$.
Choosing Cartesian coordinates $x^{a}$, the vielbein and the spin connection
may be written as $e^{a}=$ d$x^{a}$ and $\omega^{ab}=0$. This vielbein can be
gauged away with the transformation $\exp\left[  \left(  x^{a}-x_{0}%
^{a}\right)  \bm{P}_{a}\right]  $.

In this way, we have found that there is nothing pathological with an space
where the vielbein is pure gauge; it is just a natural consequence of the fact
that now the vielbein is part of a gauge connection.


\begin{thebibliography}{99}                                                                                               %
\bibitem {Zum85}B.~Zumino, \textit{Gravity Theories in More Than Four
Dimensions}. Phys. Rept. \textbf{137} (1986) 109.

\bibitem {Zwi85}B.~Zwiebach, \textit{Curvature Squared Terms and String
Theories}. Phys. Lett. B \textbf{156} (1985) 315.

\bibitem {Lov70}D.~Lovelock, \textit{The Einstein Tensor and its
Generalizations}. J. Math. Phys. \textbf{12} (1971) 498.

\bibitem {Lan38}C.~Lanczos, \textit{A Remarkable Property of the
Riemann--Christoffel Tensor in Four Dimensions}. Ann. Math. \textbf{39} (1938) 842.

\bibitem {Des05}S.~Deser, \textit{Birkhoff for Lovelock Redux}. Class. Quantum
Grav. \textbf{22} (2005) L103. arXiv: gr-qc/0506014.

\bibitem {Zeg05}R.~Zegers, \textit{Birkhoff's Theorem in Lovelock Gravity}. J.
Math. Phys. \textbf{46} (2005) 072502. arXiv: gr-qc/0505016.

\bibitem {Hen05}S.~Cnockaert, M.~Henneaux, \textit{Lovelock terms and BRST
Cohomology}. Class. Quantum Grav. \textbf{22} (2005) 2797. arXiv: hep-th/0504169.

\bibitem {Wil04}S.~Willison, \textit{Intersecting Hypersurfaces and Lovelock
Gravity}. Ph.D. Thesis, arXiv: gr-qc/0502089.

\bibitem {Aie05}M.~Aiello, R.~Ferraro, G.~Giribet, \textit{Hoffmann--Infeld
Black-Hole Solutions in Lovelock Gravity}. Class. Quantum Grav. \textbf{22}
(2005) 2579. arXiv: gr-qc/0502069.

\bibitem {Iza04}F.~Izaurieta, E.~Rodr\'{\i}guez, P.~Salgado,
\textit{Euler--Chern--Simons Gravity from Lovelock--Born--Infeld Gravity}.
Phys. Lett. B \textbf{586} (2004) 397. arXiv: hep-th/0402208.

\bibitem {All03}G.~Allemandi, M.~Francaviglia, M.~Raiteri, \textit{Charges and
Energy in Chern--Simons Theories and Lovelock Gravity}. Class. Quantum Grav.
\textbf{20} (2003) 5103. arXiv: gr-qc/0308019.

\bibitem {Sal03}P.~Salgado, F.~Izaurieta, E.~Rodr\'{\i}guez, \textit{Higher
Dimensional Gravity Invariant Under the AdS Group}. Phys. Lett. B \textbf{574}
(2003) 283. arXiv: hep-th/0305180.

\bibitem {Aro00}R.~Aros, R.~Troncoso, J.~Zanelli, \textit{Black Holes with
Topologically Nontrivial AdS Asymptotics}. Phys. Rev. D \textbf{63} (2001)
084015. arXiv: hep-th/0011097.

\bibitem {Cri00}J.~Cris\'{o}stomo, R.~Troncoso, J.~Zanelli, \textit{Black Hole
Scan}. Phys. Rev. D \textbf{62} (2000) 084013. arXiv: hep-th/0003271.

\bibitem {Cha89}A.~H.~Chamseddine, \textit{Topological Gauge Theory of Gravity
in Five and All Odd Dimensions}. Phys. Lett. B \textbf{233} (1989) 291.

\bibitem {Cha90}A.~H.~Chamseddine, \textit{Topological Gravity and
Supergravity in Various Dimensions}. Nucl. Phys. B \textbf{346} (1990) 213.

\bibitem {Ban93}M.~Ba\~{n}ados, C.~Teitelboim, J.~Zanelli,
\textit{Dimensionally Continued Black Holes}. Phys. Rev. D \textbf{49} (1994)
975. arXiv: gr-qc/9307033.

\bibitem {Tro99}R.~Troncoso, J.~Zanelli, \textit{Higher Dimensional Gravity,
Propagating Torsion and AdS Gauge Invariance}. Class. Quantum Grav.
\textbf{17} (2000) 4451. arXiv: hep-th/9907109.

\bibitem {Zan99}R.~Aros, M.~Contreras, R.~Olea, R.~Troncoso, J.~Zanelli,
\textit{Conserved Charges for Gravity With Locally AdS Asymptotics}. Phys.
Rev. Lett. \textbf{84} (2000) 1647. arXiv: gr-qc/9909015.

\bibitem {Zan00}R.~Aros, M.~Contreras, R.~Olea, R.~Troncoso, J.~Zanelli,
\textit{Conserved Charges for Even Dimensional Asymptotically AdS Gravity
Theories}. Phys. Rev. D \textbf{62} (2000) 044002. arXiv: hep-th/9912045.

\bibitem {Mor04}P.~Mora, R.~Olea, R.~Troncoso, J.~Zanelli, \textit{Finite
Action Principle for Chern--Simons AdS Gravity}. JHEP \textbf{0406} (2004)
036. arXiv: hep-th/0405267.

\bibitem {Ole04}P.~Mora, R.~Olea, R.~Troncoso, J.~Zanelli, \textit{Vacuum
Energy in Odd-Dimensional AdS Gravity}. arXiv: hep-th/0412046.

\bibitem {Nak03}M.~Nakahara, \textit{Geometry, Topology and Physics}.
Institute of Physics Publishing; 2nd edition (2003).

\bibitem {Azc95}J.~A.~de Azc\'{a}rraga, J.~M.~Izquierdo, \textit{Lie Groups,
Lie Algebras, Cohomology and some Applications in Physics}. Cambridge
University Press (1995).

\bibitem {Has03}M.~Hassa\"{\i}ne, R.~Troncoso, J.~Zanelli,
\textit{Poincar\'{e} Invariant Gravity with Local Supersymmetry as a Gauge
Theory for the M-Algebra}. Phys. Lett. B \textbf{596} (2004) 132. arXiv: hep-th/0306258.

\bibitem {Has05}M.~Hassa\"{\i}ne, R.~Troncoso, J.~Zanelli, \textit{11D
Supergravity as a Gauge Theory for the M-Algebra}. Proc. Sci. WC 2004, 006
(2005). arXiv: hep-th/0503220.

\bibitem {Man85}J.~Ma\~{n}es, R.~Stora, B.~Zumino, \textit{Algebraic Study of
Chiral Anomalies}. Commun. Math. Phys. \textbf{102} (1985) 157.

\bibitem {Tro96}M.~Ba\~{n}ados, R.~Troncoso, J.~Zanelli,
\textit{Higher-dimensional Chern--Simons Supergravity}. Phys. Rev. D
\textbf{54} (1996) 2605. arXiv: gr-qc/9601003.

\bibitem {Tro98}R.~Troncoso, J.~Zanelli, \textit{Gauge Supergravities for All
Odd Dimensions}. Int. J. Theor. Phys. \textbf{38} (1999) 1181. arXiv: hep-th/9807029.

\bibitem {Zan00b}J.~Zanelli, \textit{Chern--Simons Gravity: From }%
$2+1$\textit{\ dimensions to }$2n+1$\textit{\ dimensions}. Braz. J. Phys.
\textbf{30}, 251 (2000). arXiv: hep-th/0010049.

\bibitem {Zan02}J.~Zanelli, \textit{(Super)gravities Beyond Four Dimensions}.
arXiv: hep-th/0206169.

\bibitem {Zan05}J.~Zanelli, \textit{Lecture Notes on Chern--Simons
(Super-)Gravities}. arXiv: hep-th/0502193.

\bibitem {Ste80}K.~S.~Stelle, P.~C.~West, \textit{Spontaneously Broken de
Sitter Symmetry and the Gravitational Holonomy Group}. Phys. Rev. D \textbf{6}
(1980) 1466.

\bibitem {Bor03}A.~Borowiec, M.~Ferraris, M.~Francaviglia, \textit{A Covariant
Formalism for Chern--Simons Gravity}. J. Phys. A \textbf{36} (2003) 2589.
arXiv: hep-th/0301146.

\bibitem {Bor05}A.~Borowiec, L.~Fatibene, M.~Ferraris, M.~Francaviglia,
\textit{Covariant Lagrangian Formulation of Chern--Simons and BF Theories}.
arXiv: hep-th/0511060.

\bibitem {MorTh}P.~Mora, \textit{Transgression Forms as Unifying Principle in
Field Theory}. Ph.D. Thesis, Universidad de la Rep\'{u}blica, Uruguay (2003).
arXiv: hep-th/0512255.

\bibitem {Mor00}P.~Mora, H.~Nishino, \textit{Fundamental Extended Objects for
Chern--Simons Supergravity}. Phys. Lett. B \textbf{482} (2000) 222. arXiv: hep-th/0002077.

\bibitem {Mor01}P.~Mora, \textit{Chern--Simons Supersymmetric Branes}. Nucl.
Phys. B \textbf{594} (2001) 229. arXiv: hep-th/0008180.
\end{thebibliography}
\end{document}